\documentclass[12pt,preprint,xdvi]{aastex}

\usepackage{natbib}

\newcommand{\be}{\begin{equation}}
\newcommand{\ee}{\end{equation}}
\newcommand\eq{eq.}
\newcommand\eqs{eqs.}
\newcommand\fig{Fig.}
\newcommand\figs{Figs.}

\def\half{\hbox{${1\over2}$}}

\def\cv{{c}_{_{\rm V}}}
\def\cvo{{c}^{\rm (fi)}_{_{\rm V}}}
\def\kb{k_{_{\rm B}}}
\def\cs{c_{\rm s}}
\def\heq{H_{\rm eq}}
\def\eion{\varepsilon_{\rm ion}}
\def\DTion{\Delta T_{\rm ion}}

\begin{document}

\title{A simple model of chromospheric evaporation and condensation driven conductively in a solar flare }

\author{D.W. Longcope}
\affil{Department of Physics, Montana State University,
  Bozeman, Montana 59717}

\keywords{Sun: flares}


\begin{abstract}
Magnetic energy released in the corona by solar flares reaches the chromosphere where it drives characteristic upflows and downflows known as evaporation and condensation.  These flows are studied here for the case where energy is transported to the chromosphere by thermal conduction.  An analytic model is used to develop relations by which the density and velocity of each flow can be predicted from coronal parameters including the flare's energy flux $F$.  These relations are explored and refined using a series of numerical investigations in which the transition region is represented by a simplified density jump.  The maximum evaporation velocity, for example, is well approximated by $v_e\simeq0.38(F/\rho_{co,0})^{1/3}$, where 
$\rho_{co,0}$ is the mass density of the pre-flare corona.  This and the other relations are found to fit simulations using more realistic models of the transition region both performed in this work, and taken from a variety of previously published investigations.  These relations offer a novel and efficient means of simulating coronal reconnection without neglecting entirely the effects of evaporation.
\end{abstract}


\section{Introduction}

Solar flares are events in which large amounts of magnetic energy, stored in the coronal field, are rapidly converted to other forms.  Many of the dramatic consequences of flares result less directly from the energy release than from the large mass of chromospheric material that is heated and driven upward in a process known as chromospheric evaporation 
\citep{Canfield1980,Antonucci1999}.  Direct signatures of this process are observed in doppler shifts of hot spectral lines 
\citep{Antonucci1983,Zarro1988,Brosius2004,Milligan2009}, but the tremendous increase in emission measure of high-temperature plasma provides equally compelling, albeit indirect, evidence \citep{Neupert1968}.

Since it was first proposed chromospheric evaporation has been investigated by numerical simulations solving one-dimensional gas-dynamic equations \citep{Nagai1980,Somov1981,Peres1982,McClymont1983}.  Some investigations introduce the flare energy as a beam of non-thermal electrons impacting the chromosphere \citep{MacNiece1984,Fisher1985,Fisher1985b,Fisher1985c}, while others use an {\em ad hoc} heat source situated at the loop top, whose energy is carried to the chromosphere by thermal conduction \citep{Cheng1983,MacNiece1986}.  In both scenarios, beam-heated and conductive, 
chromospheric material is heated and driven upward at speeds comparable to, or exceeding, the local sound speed.  In many simulations there is a downward flow, called {\em chromospheric condensation}, for which some spectroscopic observations provide separate evidence \citep{Ichimoto1984,Brosius2004,Brosius2009b,Milligan2009}.

It is generally believed that magnetic reconnection is ultimately responsible for releasing the coronal magnetic energy and thereby initiating a solar flare.  The {\em ad hoc} energization invoked in one-dimensional gas-dynamic simulations mentioned above, either thermal or non-thermal, is thus intended to model reconnection.  Emission of hard X-rays or microwaves from loop footpoints provide evidence of non-thermal electrons energizing the chromosphere.  There are, however, flares in which substantial evaporation occurs in the absence of footpoint emission \citep{Zarro1988,Longcope2010}.  In such cases the downward transport of reconnection energy can be attributed to thermal conduction.  Moreover, even in those flares where it does occur, footpoint emission is observed to abate well before the flare's energy release ends.  Evaporation in the later stages of these flares is presumed to be driven by thermal conduction from the reconnection site at the loop tops.  It has been proposed that the increased density resulting from chromospheric evaporation comes to inhibit the propagation of non-thermal beams, thereby causing the transition to conduction-driven evaporation \citep{Liu2006}.

At present we can hope to conduct theoretical studies of reconnection and evaporation together 
only for cases of {\em conduction-driven} evaporation.  While theoretical models exists for non-thermal electron energization by turbulence \citep{Benz1987,Hamilton1992,Miller1996,Park1997} and by steady electric fields \citep{Litvinenko1996}, no such model includes the process of magnetic reconnection self-consistently.  Instead, most large-scale models of flare-related magnetic reconnection have been formulated using resistive MHD: fluid equations lacking a non-thermal electron population \citep{Forbes1983,Mikic1988,Amari1996,Magara1996,Nishida2009,Birn2009}.  In fluid models of fast reconnection, magnetic energy is ultimately converted to kinetic energy of flows whose shocks produce heat \citep{Petschek1964,Soward1982b,Longcope2009}, which is then is carried by thermal conduction to the chromosphere where it drives evaporation \citep{Forbes1989,Tsuneta1996}.  Several investigations have succeeded in accommodating all these effects in a single, self-consistent simulation \citep{Yokoyama1997,Yokoyama1998,Chen1999b,Chen1999}.

Due to the numerical difficulties of resolving both the coronal reconnection, the solar transition region and the field-align thermal conduction between them, few investigations have followed those of \citet{Yokoyama1997} or \citet{Chen1999} to simulate reconnection and evaporation together. Instead, the vast majority of theoretical investigations study  magnetic reconnection in isolation, omitting the process of evaporation.  This is primarily done to achieve better resolution of the coronal physics of reconnection.  There is some evidence, however, that electron acceleration, and thus presumably reconnection, occurs at densities significantly enhanced by evaporation \citep{Veronig2004,Jiang2006,Guo2012} and even continuing to increase by ongoing evaporation \citep{Liu2006}. Thus it seems unrealistic to consider either process, reconnection or evaporation, in isolation.  Moreover, many of the observational signatures with which reconnection models must ultimately make contact, such as X-ray and EUV light-curves or spectra, are direct effects of evaporation and only indirect effects of reconnection.

In place of a full, direct numerical simulation of both processes together, the effects of evaporation on coronal reconnection could be investigated using flows imposed in an otherwise coronal simulation.  Doing so would, however, require a simplified relation between reconnection energy release and characteristics of the evaporation, such as density and flow velocity.  Previous numerical studies of evaporation were generally not aimed at developing such a relation; fewer still sought one for conductively-driven evaporation.  Among the few investigations of this kind, \citet{Fisher1987} and \citet{Brown1989} each offer different analytic models for evaporation from intense beam deposition.  \citet{Fisher1989b} developed an analytic model for chromospheric condensation driven by either beam or conductive energy deposition.  \citet{Fisher1984} presents a quantitative upper bound for the evaporation velocity in terms of evaporation temperature.  The data provided in that work showed, however, actual evaporation speeds falling below this upper bound by as much as an order of magnitude.  In fact, the data presented by \citet{Fisher1984} suggest a more useful relation might exist between evaporation velocity and flare energy flux for which they offer no explanation.

The present work takes up the challenge of finding a simple relationship between flare energy release and the evaporation and condensation flows it generates.  It adopts a purely fluid model, with energy transported by thermal conduction rather than energy beams.  This is done firstly because, as explained above, fluid models remain the only kind that can provide a complete picture of flares from energy storage, to release, and to evaporation.  A second advantage is to reduce the size of available parameter space.  Non-thermal beams are characterized by an energy flux, a spectral index, and a lower-energy cut-off, and evaporation has been found to depend on all of these parameters \citep{Fisher1989b}.  We show here that thermal energy transport is effectively characterized by a single parameter, the energy flux $F$.  This would in turn be determined by the amplitude (i.e.\ Mach number) of the shocks generating the heat, which depends in turn on the current sheet at which reconnection occurs \citep{Longcope2009,Longcope2010}.  The result of the present investigation is a set of simple relationships between $F$ and the density and velocity of both chromospheric evaporation and chromospheric condensation.  These relationships provide a very simple means of using spectroscopic observations of emission from the transition region and chromosphere, to infer properties of magnetic reconnection in solar flares.

In order to facilitate the variation of parameters this investigation uses a simplified model of the pre-flare transition region.  This is described, along with the governing dynamical equations, in the next section.  Section 3 describes a typical numerical solution to these equations.  This motivates a simplified analytical model, presented in the same section, and then compared to the numerical solution in detail.  Section \ref{sec:scaling} present a large number of different numerical solutions spanning parameters.  The analytical model is used to propose scaling relations which can be fit to the runs.  Section \ref{sec:5} then applies these same scaling laws to simulations with more realistic transition region and chromosphere treatments, including results from the literature.  In nearly every case these results are found to conform reasonably well to the simple scaling laws.

\section{The Model}

In this work, as in many previous investigations \citep{Nagai1980,Somov1981,Peres1982,MacNiece1984}, evaporation is studied using one-dimensional gas-dynamic equations for evolution of the plasma along a static flux tube, parameterized by length $\ell$.  Equations for mass, momentum and energy conservation in the flux tube of uniform cross section are
\begin{eqnarray}
  {\partial\rho\over\partial t} + {\partial\over\partial\ell}(\, \rho v\, ) &=& 0  ~~, \label{eq:cont} \\
  \rho\left( {\partial v\over\partial t} + v{\partial v\over\partial\ell}\right) &=& -{\partial p\over\partial\ell} ~+~ 
  {\partial\over\partial\ell}\left( \hbox{${4\over3}$}\mu\, {\partial v\over \partial\ell} \right) ~+~ \rho g_{\parallel}~~, 
  \label{eq:mom} \\
  \cv \rho \left( {\partial T\over\partial t} +v{\partial T\over\partial\ell} \right) &=&
  - p\,{\partial v\over\partial \ell} ~+~ \hbox{${4\over3}$}\mu \left({\partial v\over \partial\ell}\right)^2
  +~ {\partial\over\partial\ell}\left( \kappa\, {\partial T\over \partial\ell} \right) +~ \dot{Q} ~~,\label{eq:erg}
\end{eqnarray}
where $\rho$ and $T$ are mass density and temperature respectively, $g_{\parallel}$ is the component of gravitational acceleration along the tube, and $\cv$ is the specific heat, described below.  The fluid moves along the flux tube at velocity $v$.  Heating and cooling due to the flare and radiative losses are included in the volumetric power $\dot{Q}$ which is described in detail in the next sections.   The total plasma pressure $p$ is found using the ideal gas equation,
\be
  p ~=~ {\kb\over\bar{m}}\, \rho\, T ~~,
\ee
where $\kb$ is Boltzmann's constant and $\bar{m}$ is the mean mass per particle.

Equations (\ref{eq:cont})--(\ref{eq:erg}) are solved within a region spanning the corona and the chromosphere, over which varying states of ionization will occur.  The focus is, however, on flare-driven flows which occur 
in plasmas heated to temperatures around or above $10^6$K, 
and for which the plasma is expected to be fully ionized.  We therefore adopt the expedient measure of solving  \eqs\ (\ref{eq:cont})--(\ref{eq:erg}) for a fully ionized plasma with coronal abundance, setting $\bar{m} = 0.593\, m_p$, where $m_p$ is the proton mass, and taking the specific heat
\be
  \cv ~=~ {3\over 2}{\kb \over \bar{m}} ~~.
  	\label{eq:cv}
\ee
The main way that ionization would affect our results would be by removing energy from the condensation shock through an additional contribution to $\cv$.  We discuss this in an appendix and show that such a contribution would have relatively minor effects on our results.

The thermal conductivity $\kappa$ depends on temperature according to the classical Spitzer form 
\citep{Braginskii1965}
\be
  \kappa ~=~ \kappa_0\, T^{5/2} ~~~~,~~~~\kappa_0 ~=~ 10^{-6} \, {\rm erg\,cm^{-1}\,s^{-1}\,K^{-7/2}} ~~.
\ee
While this classical form can become inapplicable for extremely large heat fluxes, we have found that it applies to cases where heating occurs self-consistently through shocks \citep{Guidoni2010,Longcope2010b}.  In order to simplify our analysis we use it for all of our computations.  The parallel component of dynamic viscosity $\mu$ has a temperature dependence identical to classical conductivity, but is lower by a factor proportional to  the Prandtl number $Pr$,
\be
  \mu ~=~ Pr\, {\kappa\over \cv} ~~.
\ee
The Prandtl number in fully-ionized plasma is $Pr=0.012$, reflecting the very small electron-to-ion mass ratio.  While it is, in some sense, much smaller than thermal conductivity, its effects are very important, especially in flares where $T$ has increased by an order of magnitude \citep{Peres1993}.  Since they thermalize bulk kinetic energy, shocks cannot be modeled without viscosity.  Nevertheless, some flare investigations in the literature do solve gas dynamic equations {\em without} viscosity \citep{Cheng1983}.  To make closer contact with that work, and to sharpen the shocks for easier identification, we perform most runs using an artificially small value $Pr=10^{-4}$.  To prevent under-resolution and maintain energy conservation we place a lower bound on the viscosity
\be
  \mu ~>~ 0.05\, \rho\, \cs \Delta\ell ~~,
\ee
where $\Delta\ell$ is the local grid spacing and $\cs$ is the local sound speed.  This bound is typically assumed in the regions of lowest $T$, such as the pre-flare chromosphere.

Equations (\ref{eq:cont})--(\ref{eq:erg}) are solved using a Lagrangian code similar to that described in \citet{Guidoni2010}.  In this new code the thermal conductivity term in \eq\ (\ref{eq:erg}) is differenced implicitly to permit larger time steps even in the face of extremely high temperatures.  In contrast to \citet{Guidoni2010}, magnetic tension has been omitted from \eq\ (\ref{eq:mom}) under the assumption that the flux tube is already in magnetostatic equilibrium.  This omission removes magnetic reconnection as an explicit energy source.  We follow, instead, the standard practice of replacing the omitted term with with an {\em ad hoc} heating, without a corresponding force.  In future work we will return to explore the effect of making this assumption compared to allowing reconnection to energize the flare loop legitimately.

\subsection{The initial condition --- a simplified transition region}

The initial condition for the model is a hydrostatic loop  of total length $L$, consisting of a coronal portion, transition regions (TRs), and chromospheres at each footpoint.  We add to the heating term, $\dot{Q}$, an {\em ad hoc} contribution, localized to the center of the loop (i.e.\ the loop top), to drive the evaporation and condensation via thermal conduction.  The aim of this study is to characterize the evaporation and condensation as responses to the loop-top energy input.  This characterization will depend, to some degree, on the pre-flare structure of the TR and chromosphere.

In equilibrium models, the structure of the TR and chromosphere is determined by the interplay between heating, thermal conduction, and radiation.  The relative magnitude of these competing effects can be estimated from the minimum value of the logarithmic differential emission measure (DEM), $\xi=n_e^2dz/d\ln T$, observed to be about $\xi_{\rm min}\sim 10^{27}\,{\rm cm}^{-5}$ at the TR temperature $T_{\rm tr}\sim10^5$ K \citep{Dere1982}.  The minimum DEM corresponds to conductive flux of
\be
  |F_c| ~\simeq~ {\kappa_0\over 4\kb^2}{p^2T_{\rm tr}^{3/2}\over\,\xi_{\rm min}} ~=~ 
  (4\times10^5\,{\rm cm^4\,erg^{-1}})\,p^2 ~~,
\ee
where $p$ is the equilibrium pressure at $T_{\rm tr}$.  Since active region loops have equilibrium pressures generally less than 
$10\,{\rm erg\,cm^{-3}}$, energy fluxes are of order $10^7\,{\rm erg\,cm^{-2}\,s^{-1}}$ or less.  Energy flux typically attributed to flares, above $10^9\,{\rm erg\,cm^{-2}\,s^{-1}}$, completely overwhelms this.   The most important factor determining the chromospheric response is therefore the distribution of mass, rather than the processes which created that mass distribution in the first place.

Motivated by the foregoing argument we wish to explore how different TR mass distributions affect the evaporative response to flare heating.  Towards this end we perform a series of experiments initialized with an artificial TR, of set thickness $\Delta_{\rm tr}$, separating perfectly uniform coronal and chromospheric plasmas \citep{Brannon2014}.  In these experiments gravitational stratification is omitted ($g_{\parallel}=0$) so the equilibrium has uniform pressure $p_0$.  Radiative losses are also omitted.  Coronal plasma at uniform temperature $T_{co,0}$ is maintained by heating functions at 
$\ell=\Delta_{\rm tr}$ and  $\ell=L-\Delta_{\rm tr}$.  Temperature drops to a pre-set chromospheric level, $T_{ch,0}$, across the TRs.  This is maintained by cooling functions, at $\ell=0$ and $\ell=L$, equal and opposite to the heating functions.  This construction is implemented by setting $\dot{Q}$ in equilibrium to
\be
  \heq(\ell) ~=~ A\Bigl[\, S(\ell-\Delta_{\rm tr}) - S(\ell) + S(L-\Delta_{\rm tr}-\ell) - S(L-\ell) \,\Bigr]
\ee
where the heating and cooling are distributed over a distance $2w$ according to the shape function
\be
  S(x) ~=~ \left\{ \begin{array}{lcl} 1 - (x/w-1)^2 &~~,~~& 0 < x < 2w \\ 0 &~~,~~& \hbox{otherwise} \end{array} \right. ~~,
\ee
centered at $x=w$ and vanishing at $x=0$.  The coefficient $A$ is chosen to obtain the desired temperature ratio, 
$T_{co,0}/T_{ch,0}=R_{\rm tr}$, across each TR.

\begin{figure}[htb]
\plotone{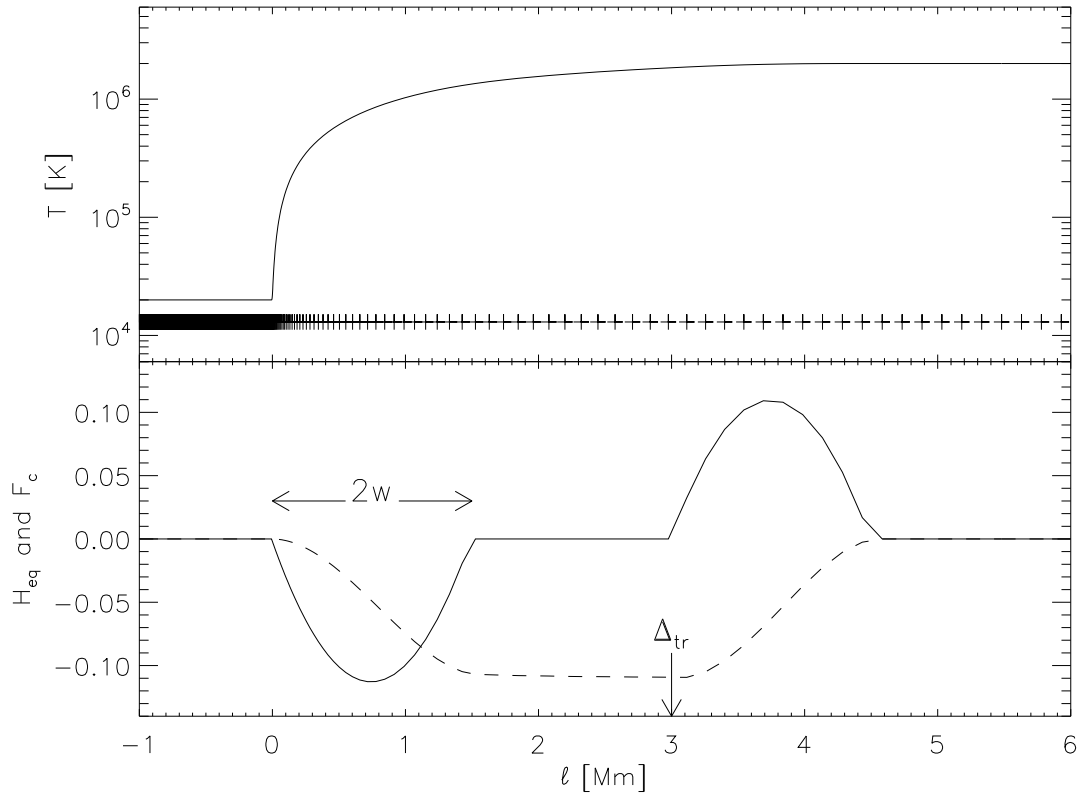}
\caption{The simplified transition region.  The top panel shows $T(\ell)$, and the bottom shows $\heq(\ell)$ 
(solid, in units of ${\rm erg\,cm^{-3}\,s^{-1}}$) and $F_c(\ell)$ (dashed, in units of $10^{8}\, {\rm erg\,cm^{-2}\,s^{-1}}$).  Plusses along the bottom of the top panel show the locations of the Lagrangian grid points.}
	\label{fig:heat}
\end{figure}

The initial condition is a static, isobaric equilibrium with $\dot{Q}=\heq(\ell)$.  From \eq\ (\ref{eq:erg}) we find the temperature distribution of the equilibrium
\be
  T^{7/2}(\ell) ~=~ T_{ch,0}^{7/2} ~-~ \hbox{${7\over2}$}\kappa_0^{-1}\, \int^{\ell}\,d\ell'\int^{\ell'}\,d\ell'' ~\heq(\ell'') ~~.
\ee
An example, shown in \fig\ \ref{fig:heat}, has $T_{co,0}=2\times10^6$ K, 
$T_{ch,0}=2\times10^4$ K, $\Delta_{\rm tr}=3$ Mm, and $w=0.75$ Mm.
The temperature changes only within the range, $0<\ell<\Delta_{\rm tr}+w=4.5$ Mm, but mostly at the outside edge of the cooling function (i.e.\ the left side or the left TR);\footnote{
Due to the artificial way we create it, we apply
the term ``transition region'' (TR) to the entire region of artificial heating and cooling, even though the majority of this range has a very shallow temperature gradient, reminiscent of the corona.}
 this point is used to define the total length of the loop.  The flux between the heating and cooling sections is $|F_c|\simeq10^7\, {\rm erg\,cm^{-2}\,s^{-1}}$, which is slightly higher than in actual TRs, but still far below the flaring energy flux.  
 This heat flux is lower than the free-streaming heat flux limit, $F_c^{\rm (fs)}={3\over2}m_en_ev_{{\rm th},e}^3$ \citep{Campbell1984} by more than an order of magnitude throughout the TR.

The Lagrangian grid points are arranged to keep roughy constant mass between them.  This has the effect of concentrating grid points in the chromosphere.  The initial locations are shown along the bottom of the top panel of \fig\ \ref{fig:heat}.

\subsection{Flare heating}

Flare energy release is modeled using a flat-topped {\em ad hoc} heating function centered at the loop top and extending a total distance of $\Delta_{\rm fl}$,
\be
  H_{\rm fl}(\ell) ~=~ \left\{ \begin{array}{lcl} {\displaystyle{F\over\Delta_{\rm fl}/2}} &~~,~~& 
  |\ell - L/2 | < \Delta_{\rm tr}/2 \\[10pt] 0 &~~,~~ & \hbox{otherwise} \end{array} \right. ~~.
  	\label{eq:Hfl}
\ee
This is added to the equilibrium heating function, $\dot{Q}=\heq+H_{\rm fl}$, to provide the term in \eq\ (\ref{eq:erg}).  That contribution is held fixed for the entire run.  

Since the solution begins with a state which is in equilibrium with heating $\heq$, the addition of $H_{\rm fl}$ has the effect of ramping the flare energy input instantaneously.  In models where flare energy is released by fast magnetic reconnection the heating occurs at a slow shock which is generated by the retraction of the reconnected flux tube \citep{Petschek1964,Soward1982}.  A particular flux tube passes the reconnection point in an instant and thereafter begins retracting \citep{Longcope2009}.  It is therefore appropriate to replace this kind of self-consistent heating with an {\em ad hoc} heating term with an instantaneous turn-on.  
Since evaporation is the focus of this work, the heating term is never turned off.

The parameter $F$ defines the total energy flux delivered to one side of the loop, and ultimately to a single footpoint.  This is one of the key parameters expected to dictate the chromospheric response.  We perform runs for a range of parameters.  The most significant variations occur for variations in energy flux $F$, loop length $L$, equilibrium values of pressure $p_0$ and TR temperature ratio $R_{\rm tr}=T_{co,0}/T_{ch,0}$.

\section{The simulations}
\label{sec:3}

\begin{figure}[htb]
\plotone{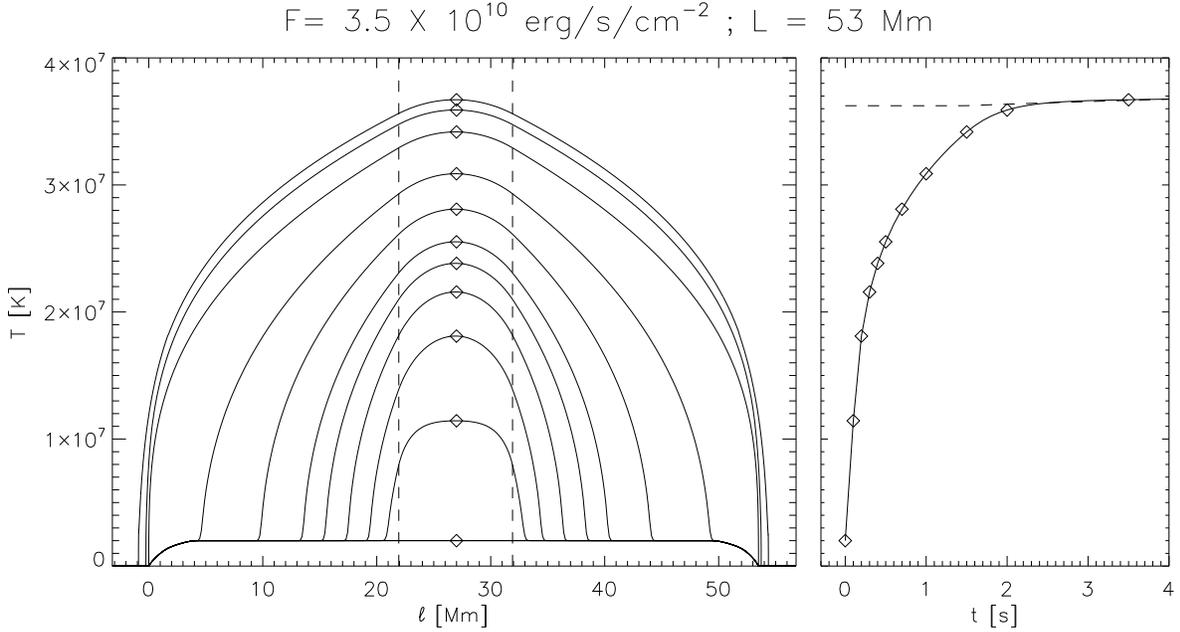}
\caption{The evolution of temperature under a flare energy flux $F=3.5\times10^{10}\,{\rm erg\, cm^{-2}\,s^{-1}}$.  The left panel shows $T(\ell)$ at a sequence of times between $t=0$ and $t=3.5$ sec.  The apex temperature, $T(L/2)$, is indicated by a diamond.  Vertical dashed lines show the region, extending $\Delta_{\rm fl}=10$ Mm, over which the flare heating is applied.  The right panel shows the continuous evolution of the apex temperature over time.  Diamonds correspond to the same times whose profiles are shown in the left panel.  The dashed curves shows the apex temperature predicted by \eq\ (\ref{eq:Tfl}).}
	\label{fig:temp_thist}
\end{figure}

Figure \ref{fig:temp_thist} shows the solution for a flux tube of length $L=53$ Mm, pressure 
$p_0=1.0\,{\rm erg\, cm^{-3}}$, subjected to a flare energy flux $F=3.5\times 10^{10}\,{\rm erg\,cm^{-2}\,s^{-1}}$, distributed over $\Delta_{\rm fl}=10$ Mm.  Its initial TR was the same one shown in \fig\ \ref{fig:heat}.  The heating produces an evolution resembling those reported in many previous simulations of this kind \citep{Cheng1983,Emslie1985}.  The temperature within the heated region (between the vertical dashed lines) rises rapidly.  At the same time thermal conduction creates fronts moving rapidly in both directions.  By $t=1.5$ s these fronts have reached the TRs and the loop top temperature slows its rise. After $t=2.0$ s the apex temperature has achieved a steady state of $T_{\rm fl}=3.7\times10^7$ K.  After that the temperature profile changes very little over the bulk of the loop.  The heat flux $F$ is being delivered to the chromosphere where it generates evaporation flow.

\begin{figure}[htb]
\plotone{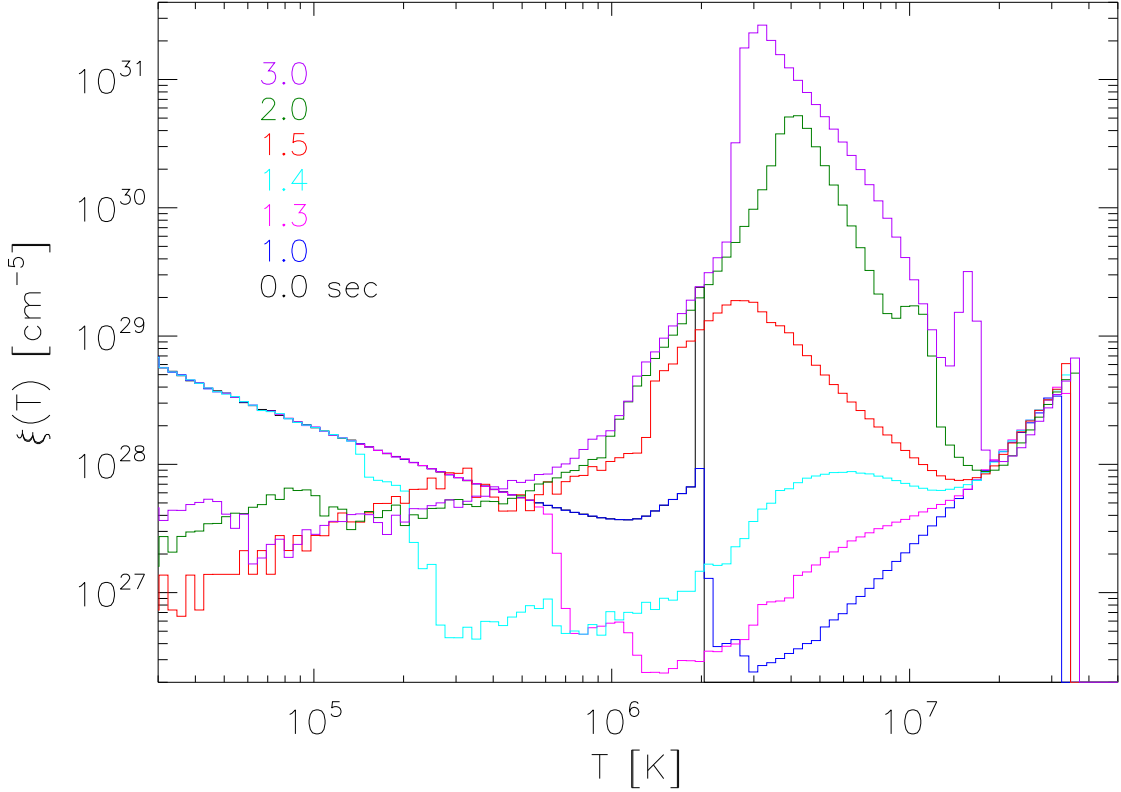}
\caption{The logarithmic DEM evolving during the initial phases shown in \fig\ \ref{fig:temp_thist}.  The curves follow a sequence progressing from low to high around $T\simeq3\times10^6$ K, matching the times listed from low to high along the left.  (Color online).}
	\label{fig:DEM}
\end{figure}

The loop's DEM undergoes a characteristic evolution during as the conduction front travels to the TR and initiates evaporation.  Figure \ref{fig:DEM} shows the logarithmic differential emission measure, $\xi(T)=n_e^2d\ell/d\ln T$, throughout the evolution depicted in \fig\ \ref{fig:temp_thist}.  The initial DEM has a sharp peak at the coronal temperature ($T=2\times10^6$ K) and a sloping TR below.  (The simplified TR creates a minimum $\min(\xi)\simeq3\times10^{27}\,{\rm cm^{-5}}$, slightly higher, and at higher temperature, than a real TR).  The conduction front shifts the emission from the coronal peak into an increasing slope up to the flare peak at $T\simeq3\times10^7$, making no change to the TR which has not yet been disturbed.  As the conduction front crosses the TR, over the times $1.3\le t\le1.5$ sec, the DEM below the corona is rapidly eroded away: a steep drop reaches $T=7\times10^5$ K and $T=2\times10^5$ K at $t=1.3$ and $1.4$ sec respectively.  This emission is piled into a peak around $3\times10^6$ K.  At later times the TR has been thinned enough that $\xi\simeq3\times10^{27}\, {\rm cm^{-5}}$, and the peak at $3\times10^6$ K grows by evaporation.

\begin{figure}[htp]
\plotone{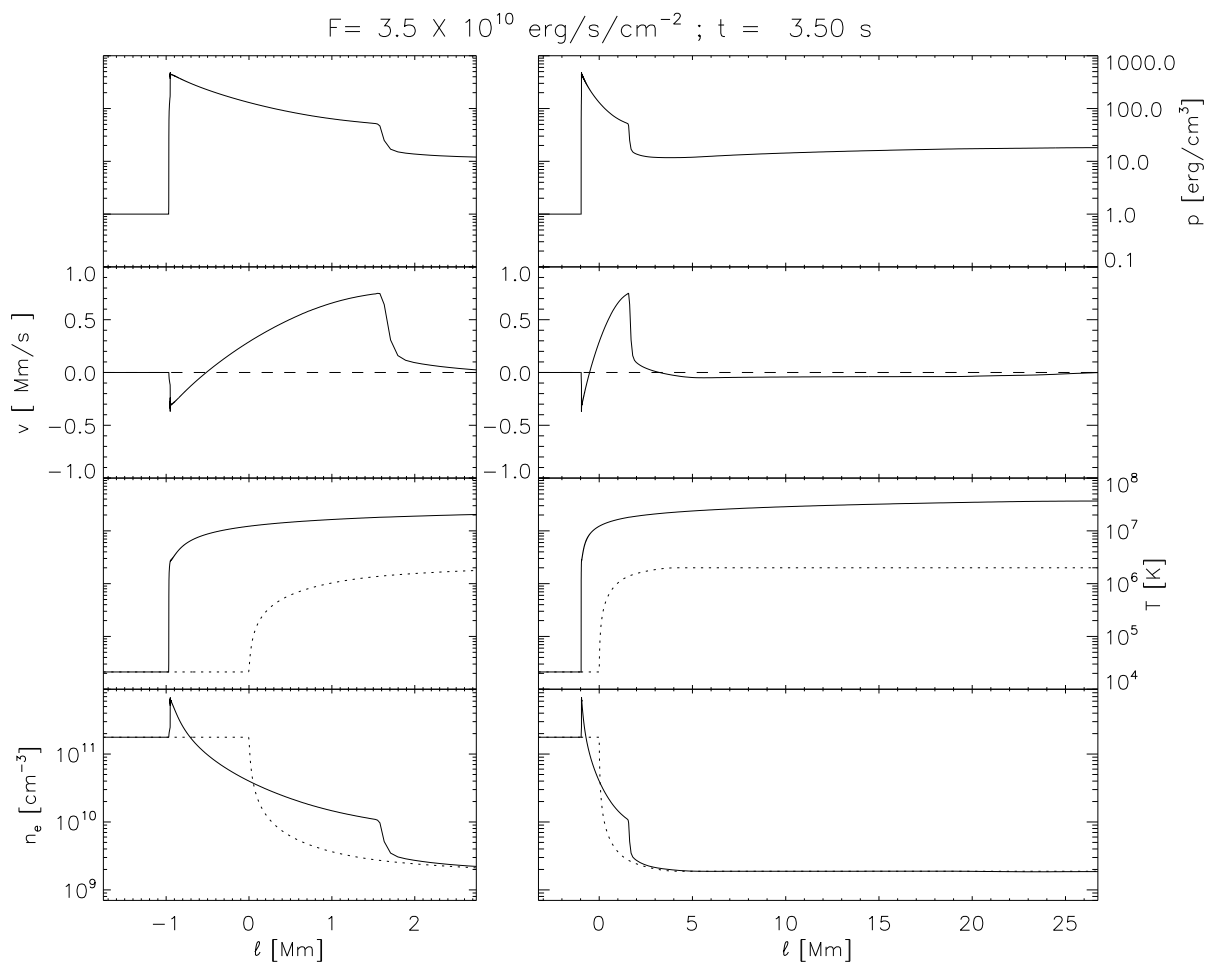}
\caption{The structure of the evaporation flow at the last time shown in \fig\ \ref{fig:temp_thist}.  The right column shows the entire left side of the flux tube, and the left column shows the region around the TR.  The rows shows, from top to bottom, pressure, velocity, temperature and electron density.  The dotted curve shows the initial profiles of temperature and density (initial pressure and velocity are both uniform).}
	\label{fig:tube_prof}
\end{figure}

The evaporation phase begins once the conduction front has reached the TR and the peak coronal temperature has plateaued: by about $t=2.0$ secs in the present case.  
Figure \ref{fig:tube_prof} shows details of this phase phase from $t=3.5$ s when the conduction front has penetrated $1.0$ Mm into the chromosphere.  
This penetration occurs at the head of the conduction front which takes the form of a downward propagating shock
of extremely high Mach number: the pressure jumps by a factor 480 and the density jump, a factor of $3.85$, is nearly 4, the maximum value permitted by Rankine Hugoniot relations.   The post-shock material is downflowing at 
$v_{c}\simeq-370$ km/sec, in a {\em chromospheric condensation} \citep{Fisher1985}; we henceforth refer to this shock as the {\em condensation shock}.  An upward propagating shock, the {\em evaporation shock}, has reached $\ell=1.5$ Mm by this time.  The density there jumps by a factor slightly greater than four and the peak velocity is $v_{e}=+750$ km/sec.  The shock is well defined because we have used the anomalous Prandtl number $Pr=10^{-4}$.  Between these two shocks lies a {\em rarefaction wave} over which the velocity changes continuously between downflow and upflow, and the density increases by a factor close to $100$.

\subsection{An analytic model of the evaporation}
\label{sec:analytic}

The foregoing simulation suggests a simple model for the evaporation.  The conduction front overtakes the TR rapidly enough that the entire region is raised to a uniform temperature, $T_*$, before any dynamical response can occur.  The TR then functions as an initial pressure discontinuity whose subsequent evolution is an isothermal Riemann problem.  Since the TR is now at uniform temperature, the pressure ratio across the jump matches the density ratio which is inversely related to the temperature ratio across the pre-flare TR,
\be
  {p_{ch}\over p_{co}} ~=~ {\rho_{ch,0}\over\rho_{co,0}} ~=~ {T_{co,0}\over T_{ch,0}} ~=~ R_{\rm tr} ~~.
\ee

Under isothermal dynamics, the initial pressure jump decomposes into a shock and a rarefaction wave \citep{Fabbro1985}.  If the initial jump is thin enough the rarefaction wave (RW) will be self-similar \citep{Landau1959,Fabbro1985}
\begin{eqnarray}
  v(\ell,t) &=& {\ell\over t} ~+~ a ~~,\\
  \rho(\ell,t) &=& \rho_0\, \exp\left( -{\ell\over at} \right) ~=~ \rho_0\, \exp\left( 1 - {v\over a} \right) ~~, \label{eq:rho_rw}
\end{eqnarray}
where $a=\sqrt{p/\rho}=\sqrt{\kb T_*/\bar{m}}$ is the iso-thermal sound speed, and $\rho_0$ is a constant determined from the surrounding solution.

\begin{figure}[htp]
\plotone{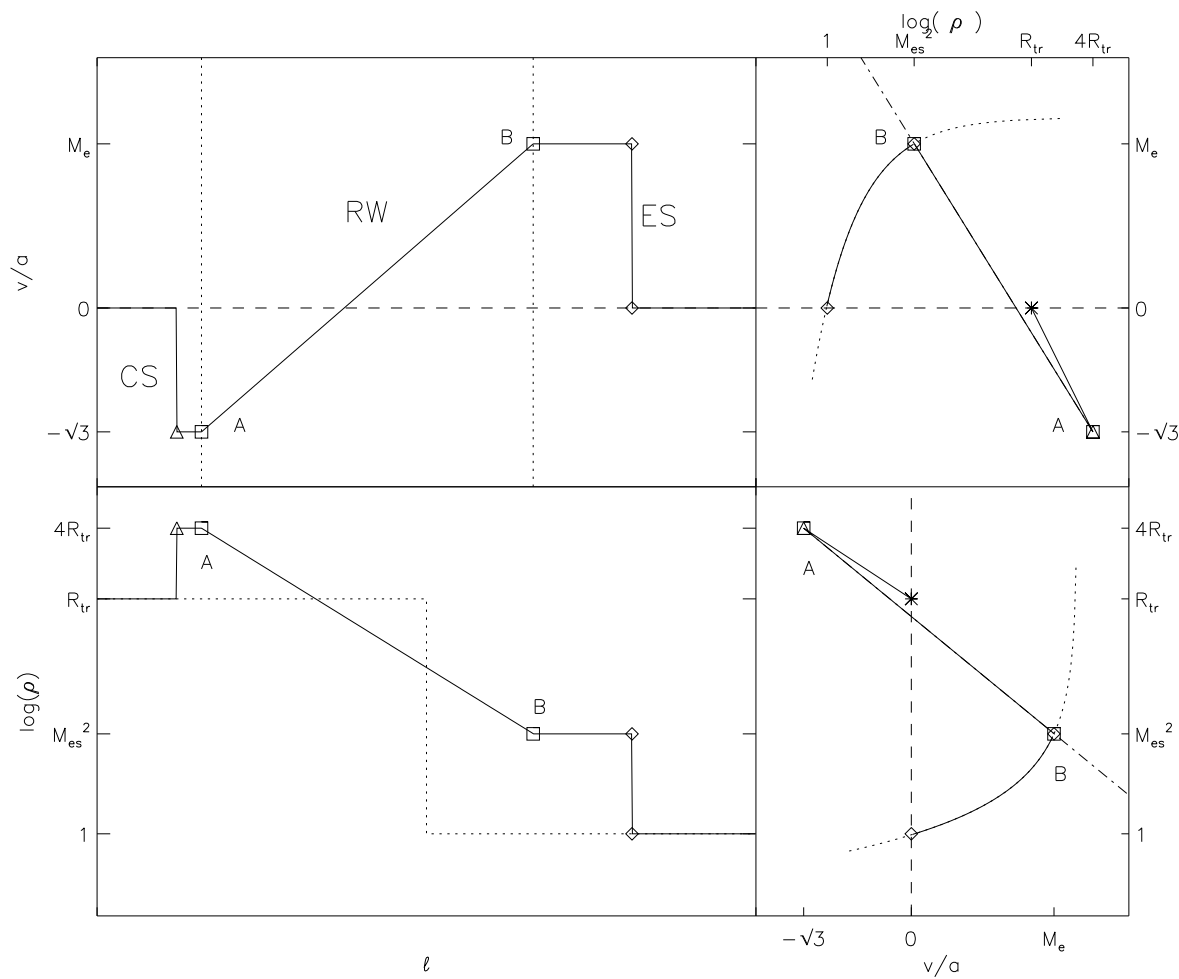}
\caption{Schematic plot of the Riemann problem solution.  The left column shows $v(\ell)/a$ (top) and $\rho(\ell)$ (bottom, on a logarithmic scale) versus position $\ell$.  The right column shows the same quantities plotted against the other (i.e.\ $v/a$ {\em vs.} $\rho$, in the upper right).  The dash-dotted line beginning at A and passing though B shows the equation for an isothermal, self-similar RW, namely \eq\ (\ref{eq:rho_rw2}).  The dotted curves show the relation for an isothermal shock.  These intersect at point B, marking the rightmost point in the RW.}
	\label{fig:Riemann}
\end{figure}

In the complete solution, illustrated in \fig\ \ref{fig:Riemann}, the left edge is the condensation shock (CS), which heats the plasma to its uniform temperature of $T_*$, with (adiabatic) sound speed $c_{s,*}=\sqrt{5/3}\,a$.  Because the CS has an extremely high Mach number it increases the density to $4\rho_{ch,0} = 4R_{\rm tr}\,\rho_{co,0}$.  The post-shock flow speed for a hypersonic shock propagating into stationary plasma is
\be
   v_{c}~=~-{3\over\sqrt{5}}\,c_{s,*} ~=~ -\sqrt{3}\, a ~~,
\ee
directed downward ($v_c<0$).  This is the condensation velocity, lying at the left hand edge of the RW and labeled A in \fig\ \ref{fig:Riemann}.  Applying these two conditions to point A allows $\rho_0$ to be eliminated from \eq\ (\ref{eq:rho_rw})
\be
  {\rho(\ell,t)\over\rho_{co,0}} ~=~ 4R_{\rm tr}\, \exp\left( -\sqrt{3} - {v\over a} \right) ~~.
  	\label{eq:rho_rw2}
\ee

The evaporation shock (ES) propagates into the stationary coronal plasma to the right.  In the reference frame of the ES, 
the coronal material is flowing leftward with isothermal 
Mach number, $M^{(it)}_{es}$.  Because it is an isothermal shock the post-shock flow, in the ES reference frame, has isothermal Mach number $1/M^{(it)}_{es}$.  In the non-moving reference frame, the evaporation flows at velocity $v_e$ to the right, while the pre-shock material is stationary.  The velocity difference between pre-shock and post-shock velocities is  independent of reference frame so
\be
  {v_e\over a}  ~=~ M^{(it)}_{es} ~-~ {1\over M^{(it)}_{es}} ~~.
\ee
The density jump across the isothermal shock is $\rho_{e}/\rho_{co,0}=[M^{(it)}_{es}]^2$.  

\begin{figure}[htp]
\epsscale{0.6}
\plotone{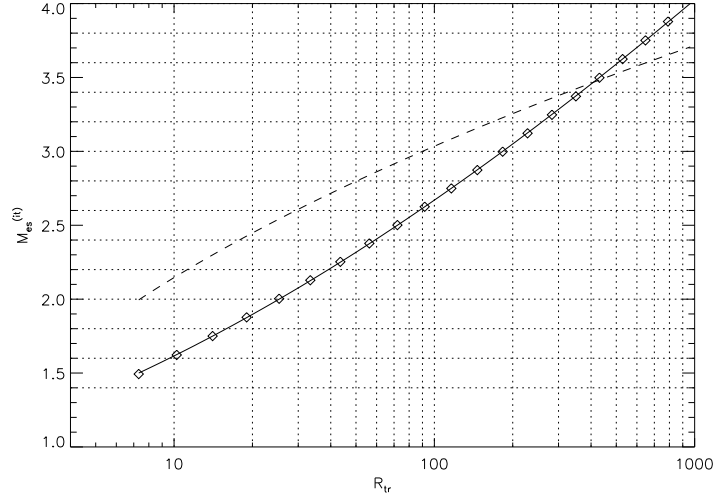}
\caption{Isothermal Mach number of the evaporation shock, $M^{(it)}_{es}$, as a function of $R_{\rm tr}$ (solid curve), satisfying \eq\ (\ref{eq:Mites}).  Diamonds follow the empirical fit \eq\ (\ref{eq:Mites_fit}).  The dashed curve shows the upper bound, \eq\ (\ref{eq:FCM}), derived by \citet{Fisher1984}.}
	\label{fig:mach}
\end{figure}

The right hand edge of the RW, labelled B in \fig\ \ref{fig:Riemann}, must match the conditions of the ES described above.  Applying these conditions to \eq\ (\ref{eq:rho_rw2}) yields the relation
\be
  \Bigl[M^{(it)}_{es}\Bigr]^2 ~=~ 4R_{\rm tr}\,\exp\left( -\sqrt{3} - M^{(it)}_{es} + {1\over  M^{(it)}_{es} } \right) ~~,
  	\label{eq:Mites}
\ee
which must be satisfied by $M^{(it)}_{es}$.  Figure \ref{fig:mach} shows the solution for a range of TR density ratios $R_{\rm tr}$.  Diamonds follow an empirical fit 
\be
  M^{(it)}_{es} ~\simeq~ 2.670 ~+~ 1.209\, \log(R_{\rm tr}/100)\,\Big[ ~1 + 0.126\, \log(R_{\rm tr}/100)~\Bigr]~~.
  	\label{eq:Mites_fit}
\ee
Due to its logarithmic dependence on $R_{\rm tr}$ the isothermal Mach number of the ES falls within the narrow range between $2.5$ and $3.5$ for most reasonable assumptions above the pre-flare transition region.

This model can be compared to an upper bound obtained by \citet{Fisher1984}.  They used the isothermal momentum equation under the assumption that the entire pressure drop matched the pre-flare values, and the kinetic energy in the condensation was ignorable.  By doing so they obtained the upper bound on the evaporation velocity,
\be
  v_{_{\rm FCM}} ~=~ \sqrt{(6/5)\,\ln R_{\rm tr}}\,c_{s,*} ~=~ \sqrt{2\ln R_{\rm tr}} \, a ~~.
  	\label{eq:FCM}
\ee
This is plotted as a dashed line in \fig\ \ref{fig:mach}.  The discrepancies are due to the different assumptions used in the two derivations.

\subsection{Modeling the simulation}
\label{sec:sim_model}

\begin{figure}[htp]
\epsscale{0.9}
\plotone{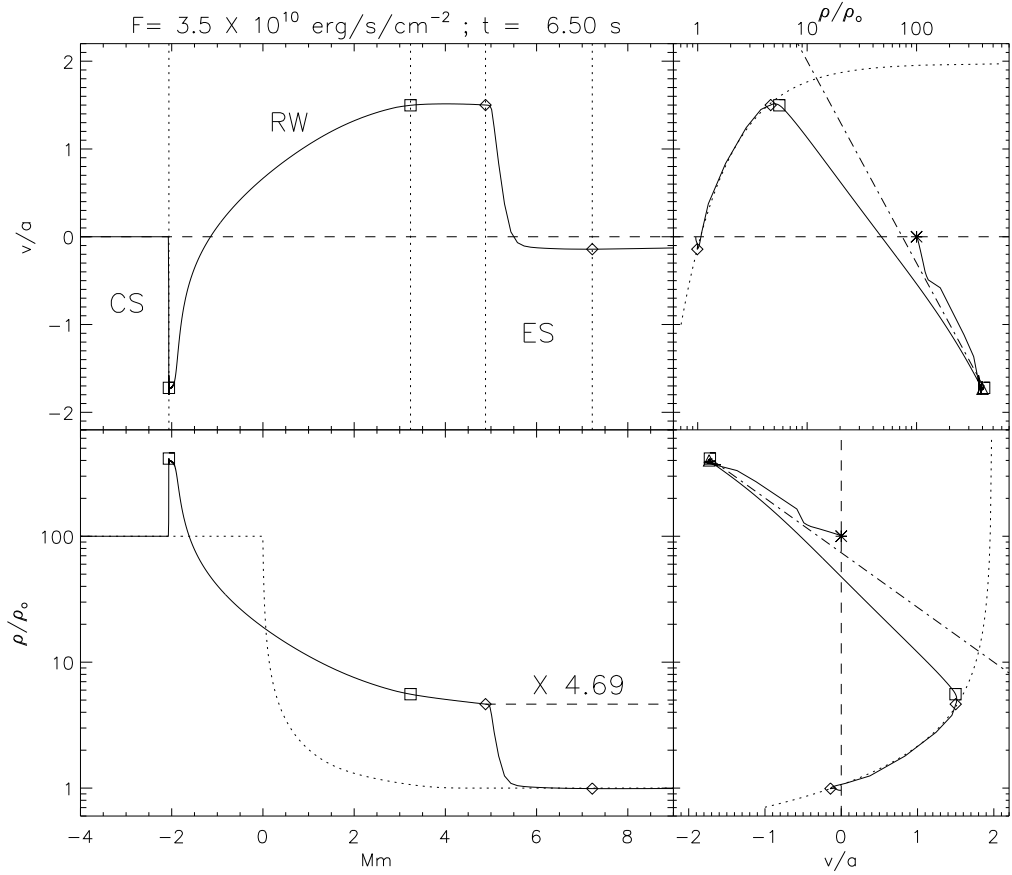}
\caption{The solutions from \fig\ \ref{fig:tube_prof}, at later time $t=6.25$ sec, plotted in the same manner as \fig\ \ref{fig:mach}.  Velocity is scaled to the local isothermal sound speed, $a$, and density to the pre-flare coronal value $\rho_{co,0}$.  Vertical dotted lines in the upper right panel show the boundaries of the RW and the ES.  The right column plots the scaled velocity against normalized density.  Dotted curves show the relation for an isothermal shock, and the dash-dotted line shows the equation for an isothermal, self-similar RW, namely \eq\ (\ref{eq:rho_rw2}).  
A triangle in the right panels shows the theoretical point in density-velocity space where a hypersonic condensation shock would fall.}
	\label{fig:struct}
\end{figure}

Figure \ref{fig:struct} shows the numerical solution from \fig\ \ref{fig:tube_prof}, at a somewhat later time ($t=6.5$ s), 
plotted in the same manner as the sketch in  \fig\ \ref{fig:Riemann}.  Because the solution includes viscosity, albeit artificially low, the ES is somewhat broadened.  The CS satisfies the structure assumed in the simplified model: 
$\rho_c=4R_{\rm tr}\rho_{co,0}$ and $v_c=-\sqrt{3}\, a$.  Its actual location, plotted with a square, lies almost on top of this theoretical one, marked with a triangle in the right column of \fig\ \ref{fig:struct}.  The ES also conforms to the assumption of an isothermal shock, marked by a dotted curve in the right column.
 
The actual RW does not, however, conform to the self-similar, isothermal structure.  The velocity is not linear with distance, nor is the density exponential.  The path between squares, in the right column of \fig\ \ref{fig:struct}, falls off the isothermal model shown by dash-dotted lines.  It is temperature variation along the RW that causes its track to fall beneath the isothermal (dash-dotted) one in the lower right.  This curve intersects the shock curve at a point of lower density and lower velocity.  Thus the prediction of the isothermal model, i.e.\ \eq\ (\ref{eq:Mites}), provides an upper bound on the isothermal Mach number of the actual shock.  For this case, with $R_{\rm tr}=100$, the isothermal model predicts $M^{(it)}_{es}=2.67$ and thus, $\rho_e/\rho_{co,0}=(2.67)^2=7.13$ and $v_e/a = 2.67-(2.67)^{-1}=2.30$.  The actual values, 4.69 and 1.50 respectively, fall below those values.  The isothermal Mach number of this shock, $M^{(it)}_{es}=\sqrt{4.69}=2.16$, lower than the prediction from \eq\ (\ref{eq:Mites}).  This is the effect of temperature variation along the RW.

\begin{figure}[htp]
\plotone{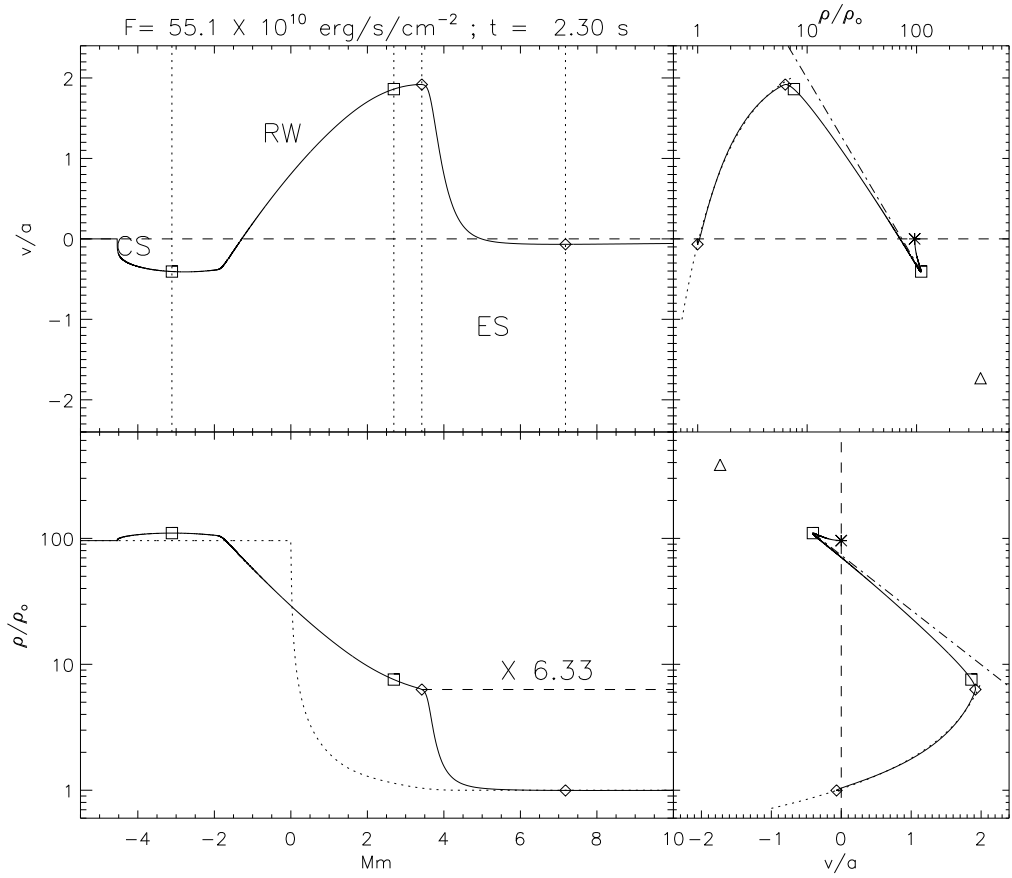}
\caption{The solutions from a run like \fig\ \ref{fig:struct} but with $F=5.5\times10^{11}\,{\rm erg\,cm^{-2}\,s^{-1}}$, plotted at 
 $t=2.3$ s in the same manner as \fig\ \ref{fig:struct}.}
	\label{fig:struct6}
\end{figure}

At higher values of heat flux $F$ the evaporation/condensation structure becomes more isothermal owing to the larger conductivity at higher temperatures.  A solution of this kind, shown in \fig\ \ref{fig:struct6}, has a RW falling closer to the isothermal model (dash-dotted curve).  At these higher fluxes, however, the RW begins to affect the CS, and the latter departs from the assumption of a hypersonic shock, as evident in the separation between square and triangle in the right column of plots.  The density behind the CS is lower than $4\rho_{ch,0}$, and the velocity greater than $-\sqrt{3}a$.  We discuss in the next section the level of energy flux required to produce this departure from the analytic model.

The analytic model breaks down for different reasons in the opposite limit of very small energy fluxes.  In this case the conduction fronts move very slowly from the loop top.  In one case (not shown) with $F=10^9\,{\rm erg\,cm^{-2}\,s^{-1}}$, the steady-evaporation phase is achieved only after $30$ s compared to 2 s, in \fig\ \ref{fig:temp_thist}.  
The loop-top heating also drives flows downward, as evident on the right column of \fig\ \ref{fig:tube_prof}.  When the conduction front arrives late, this flow impinges on the evaporation flow.  The result is a departure from the analytic model of \fig\ \ref{fig:Riemann}, and a systematically low evaporation flow.

\begin{figure}[htp]
\plotone{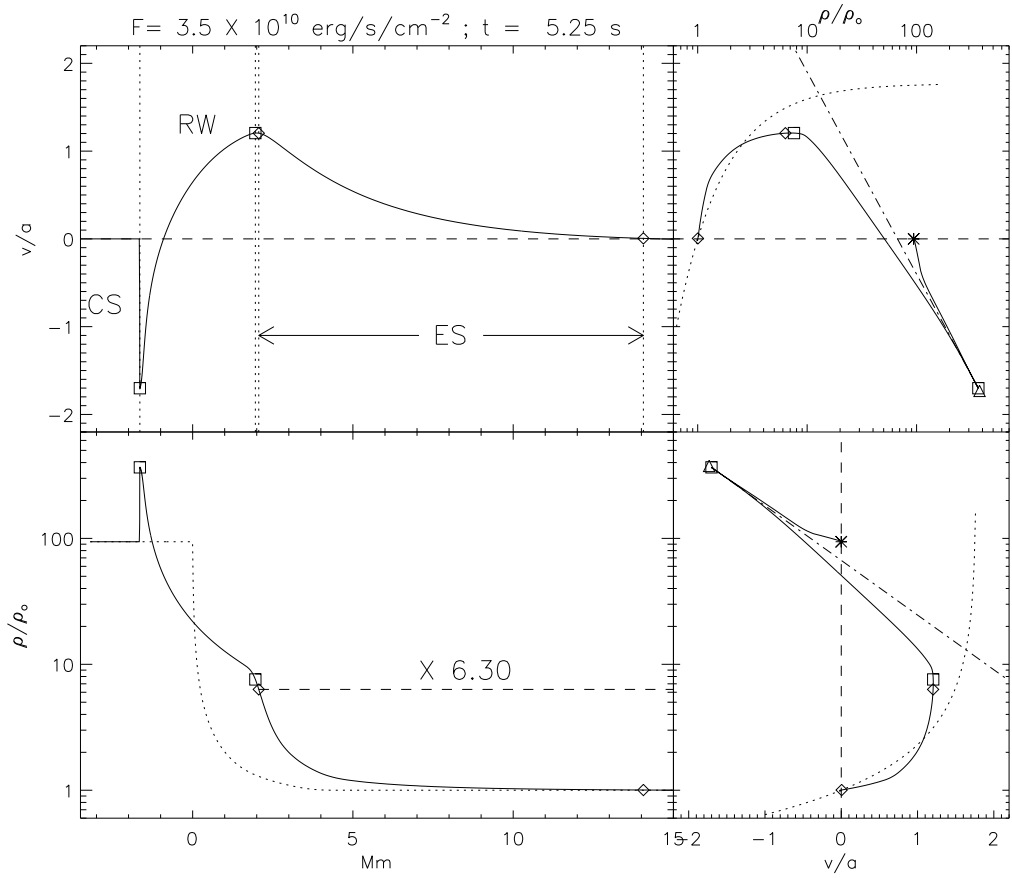}
\caption{The solutions from a run like \fig\ \ref{fig:struct}, but with physical viscosity, $Pr=0.012$, plotted in the same manner as \fig\ \ref{fig:struct}.}
	\label{fig:struct5}
\end{figure}

Using the physical viscosity, characterized by $Pr=0.012$, provides a more realistic, but less clear picture of the evaporation process.  Figure \ref{fig:struct5} shows the results of making this change to the simulation discussed above.  The larger viscosity spreads the ES considerably, causing it to overlap with the RW.  As a result, the phase space curve of the ES deviates from the isothermal curve (dotted) by veering toward the RW curve (dash-dotted).  The point of maximum velocity is adopted at the point separating these two structures, now largely merged.  This occurs at a lower velocity ($v_e/a=1.20$) and higher density ($\rho_e/\rho_{co,0}=6.30$) than in the case with a well-defined ES.  In addition, the ES continues expanding, causing these values to change with time.  We therefore continue using the anomalously low value, $Pr=10^{-4}$ to obtain clear values.  We recognize that these values will differ from those with physical viscosity in the manner just described.

\section{Scaling of the evaporation}
\label{sec:scaling}

\subsection{Variation in flare and loop parameters}

We next explore how the structure of the evaporation and condensation varies with parameters.  Twenty-eight different simulations are performed with different values of $L$, $F$, $p_0$ and $R_{\rm tr}$.  Loop lengths range from $9$ to $86$ Mm, fluxes from $10^{9}$ to $10^{12}\,{\rm erg\,cm^{-2}\,s^{-1}}$, initial pressures from $0.3$ to $3.0\,{\rm erg\,cm^{-3}}$, and $R_{\rm tr}$ from 40 to 100.  For simplicity we hold fixed the TR structure by keeping the values  
$\Delta_{\rm tr}=3$ Mm, $w=750$ km; the chromosphere is kept at $T_{ch,0}=20,000$ K, but the corona is at 
$T_{co,0}=R_{\rm tr}T_{ch,0}$ which varies as $R_{\rm tr}$ is varied.   The heating profile was also kept fixed with $\Delta_{\rm fl}=10$ Mm.  Each simulation was run past the time the apex temperature plateaued at $T_{\rm fl}$.  At that point the shocks, ES and CS, were identified and characterized.

\begin{figure}[htb]
\epsscale{1.0}
\plotone{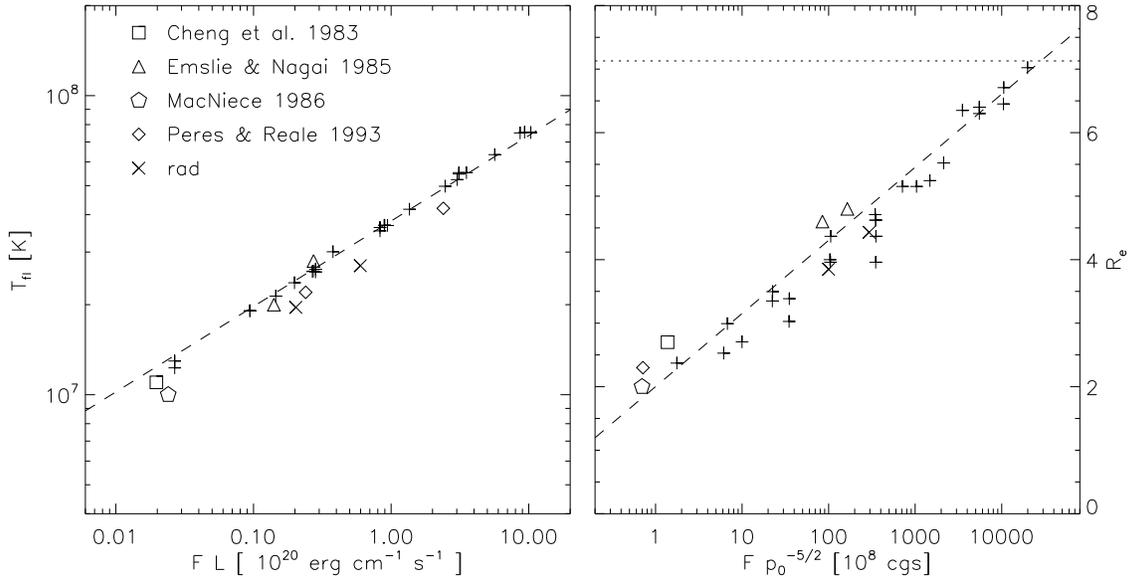}
\caption{The flare temperature $T_{\rm fl}$ (left) and ES density enhancement ratio $R_{es}$ (right) of flare simulations.  The plusses are the runs with artificial TR described in Sec.\ \ref{sec:scaling}.  Other symbols show runs with more realistic TR treatments from the literature and  this work, as described Sec.\ \ref{sec:5}.  The left panel shows $T_{\rm fl}$, in Kelvin, {\em vs.} 
$FL$ (in units of $10^{20}\, {\rm erg\,cm^{-1}\,s^{-1}}$), with a dashed line showing \eq\ (\ref{eq:Tfl}).  The right panel shows $R_{es}$ {\em vs.} $F\,p_0^{-5/2}$ (in units of $10^8\,{\rm erg^{-3/2}\,cm^{-9/2}\,s^{-1}}$), and the dashed line shows relation (\ref{eq:Res}).}
	\label{fig:temp_fit}
\end{figure}

The peak apex temperature, $T_{\rm fl}$, is achieved when the flare energy flux balances the power driving the evaporation.  
The heat flux reaching the evaporating plasma, at lower temperature, will be
$\sim \kappa_0T_{\rm fl}^{7/2}/L$.  Equating this with the input flux yields an expected scaling
\be
  T_{\rm fl} ~=~ C_T\, (FL/\kappa_0)^{2/7} ~~,
  	\label{eq:Tfl}
\ee
where $C_T$ is a dimensionless constant.  The same form is given in the appendix of \citet{Fisher1989b}.
The left panel of \fig\ \ref{fig:temp_fit} shows $T_{\rm fl}$ for the 28 runs (plusses) {\em vs.} the product $FL$, where $L$ is the {\rm full} loop length.  The dashed line shows \eq\ (\ref{eq:Tfl}) with $C_T=1.46$, 
found from a fit to all 28 runs.

According to the analytic model of Sec.\ \ref{sec:analytic}, the isothermal Mach number of the ES, $M^{(it)}_{es}$, depends only on the pre-flare chromospheric/coronal density ratio $R_{\rm tr}$ (and on that only logarithmically).  Application to an actual solution showed, however, that temperature gradient within the RW led to a slightly lower value of $M^{(it)}_{es}$.  For the cases of small viscosity, however, the ES is a simple isothermal shock and therefore its density enhancement is
\be
  R_{es} ~=~{\rho_e\over\rho_{co,0}} ~=~ \Bigl[\, M^{(it)}_{es} \, \Bigr]^2 ~~.
\ee
This will have an upper bound dependent on $R_{\rm tr}$, and a value depending on other factors responsible for the temperature variation within the RW.  Figure \ref{fig:temp_fit} shows that the observed ES density enhancement, $R_{es}$, 
is ordered by the quantity, $F/p_0^{5/2}$, identified using multivariate regression 
to the 28 plusses.   An empirical relation,
\be
  R_{es} ~=~ \hbox{${1\over 2}$}\, \ln(F\,p_0^{-5/2}/C_M) ~+~ 7.13 ~~,
  	\label{eq:Res}
\ee
lies near, and slightly above, much of the data, where $C_M=2.8\times10^{12}\,{\rm erg^{-3/2}\,cm^{-9/2}\,s^{-1}}$. Expression (\ref{eq:Mites_fit}) predicts that the upper bound to the density enhancement of, $R_{es}=(2.67)^2=7.13$. The empirical fit, \eq\ (\ref{eq:Res}), would exceed the maximum for $F\,p_0^{-5/2}>C_M$.  This value is above the point where the analytic model becomes untenable due to the reduction in Mach number of the CS, as shown in \fig\ \ref{fig:struct6}.

\begin{figure}[htb]
\epsscale{0.9}
\plotone{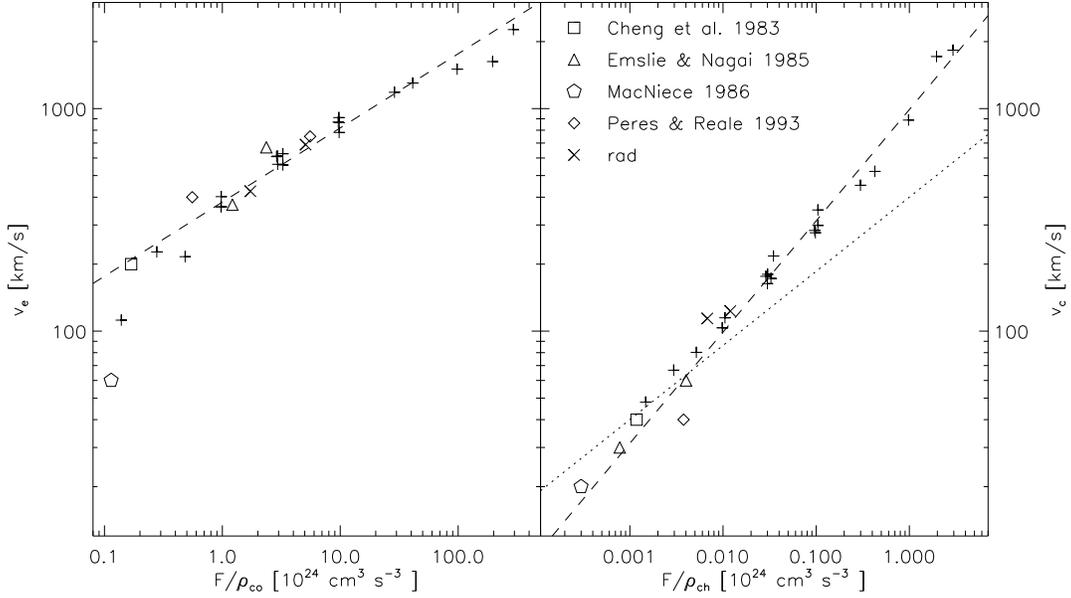}
\caption{The velocity of the evaporation (left) and and condensation (right) from flare simulations, plotted in km/s.  The plusses are the runs with artificial TR described in Sec.\ \ref{sec:scaling}.  Other symbols show runs with more realistic TR treatments from the literature and this work, as described in Sec.\ \ref{sec:5}.  The left panel shows $v_e$ {\em vs.} 
$F/\rho_{co,0}$ (in $10^{24}\,{\rm cm^{3}\,s^{-3}}$), with a dashed line showing \eq\ (\ref{eq:ve}).  
The right panel shows $v_e$ {\em vs.} $F/\rho_{ch,0}$ (in $10^{24}\,{\rm cm^{3}\,s^{-3}}$), with a 
dashed line showing \eq\ (\ref{eq:vc}).  The dotted line shows the scaling derived by \citet{Fisher1989b}, and reproduced here 
as \eq\ (\ref{eq:F89}).}
	\label{fig:vel_fit}
\end{figure}

The analytic solution shown in \fig\ \ref{fig:Riemann} has a total kinetic energy (per unit area) 
\be
  E_K ~=~ \int\, \half \rho v^2\, d\ell ~=~ \half \rho_{co,0}\, a^3\, t\, \int {\rho(\tilde{\ell})\over \rho_{co,0}}\, 
  \Bigl[M^{(it)}(\tilde{\ell})\Bigr]^2d\tilde{\ell} ~~,
  	\label{eq:EK}
\ee
where $\tilde{\ell} = \ell/at$ is the similarity variable of the solution, and $M^{(it)}=v/a$ is the scaled velocity plotted in the figure.  The integral in the final expression depends on the scaled analytic solution and therefore on $M^{(it)}_{es}$ only.  Assuming the flare energy flux, $F$, exactly balances this energy requirement (i.e. $F=dE_K/dt$), the isothermal sound speed within  the evaporation flow is 
\be
  a~\sim~\left({F/\rho_{co,0}}\right)^{1/3}~~.
  	\label{eq:a_evap}
\ee
 If we neglect the logarithmic dependence of $M^{(it)}_{es}$, 
the evaporation flow speed will scale in the same manner
\be
  v_e ~=~ C_e\,(F/\rho_{co,0})^{1/3} ~~.
  	\label{eq:ve}
\ee
The best fit to the 28 plusses, 
shown in \fig\ \ref{fig:vel_fit}, is for $C_e=0.38$.  The points at the very left and very right of the range trend below this fit.  The explanation may be the departure from the analytic model at high and low fluxes, described briefly in Sec.\ \ref{sec:sim_model}.

\citet{Fisher1989b} estimates  the height of the pressure peak separating evaporation from condensation 
to scale as $p_{e}\sim(F^2\rho_{co,0})^{1/3}$ \citep[see also][]{Fisher2012}.  While he does not go on to estimate the evaporation flow speed we can take the additional step of identifying the sound speed within the evaporation as $a\sim \sqrt{p_{e}/\rho_{co,0}}$ to obtain a scaling identical to (\ref{eq:a_evap}).  While Fisher's logic differs from that following \eq\ (\ref{eq:EK}) above, both produce the same scaling, as would simple dimensional analysis \citep{Fisher1989b}.

It is noteworthy that we found condensation shocks in every run, including those of the smallest flare energy fluxes.  Thermal conduction therefore differs from non-thermal deposition, which generates condensation only when the flux exceeds a threshold $F_{\rm thr}$.  This threshold is found to be between $F_{\rm thr} = 3\times10^{9}$ and
$3\times10^{11}\,{\rm erg\,cm^{-2}\,s^{-1}}$ depending on pre-flare pressure and the beam's spectral properties \citep{Fisher1989b}.    Non-thermal energy fluxes below the threshold produce {\em gentle evaporation}, in which there is little observable condensation.  Fluxes above the threshold drive {\em explosive evaporation}, accompanied by chromospheric condensation.  We find that thermal conduction always leads to explosive evaporation; a fact that has been noted before \citep{Fisher1989b}.  

If the isothermal model of Sec.\ \ref{sec:analytic} applied all the way to the CS, then the condensation velocity would scale as $v_c\sim a \sim (F/\rho_{co,0})^{1/3}$, just like the evaporation velocity.  The significant temperature variation within the RW, however, invalidates this conclusion.  Instead we find, using multivariate regression 
to the 28 plusses, an empirical scaling
\be
  v_c ~=~ C_c\,(F/\rho_{ch,0})^{1/2} ~~,
  	\label{eq:vc}
\ee
where $C_c=10^{-4}\,{\rm cm^{-1/2}\,s^{1/2}}$, as shown by the dashed line in the right panel of  \fig\ \ref{fig:vel_fit}.

\citet{Fisher1989b} used a different analytical approach to find a comparable scaling for condensation velocity.  For the case of thermal conduction or beam heating lower than the explosive threshold (so-called gentle evaporation) he finds condensation velocity (given in his \eq\ [34a])
\be
  v^{\rm (F89)}_{c}~\simeq~ 0.4\left({F/\rho_{ch,0}}\right)^{1/3} ~~,
  	\label{eq:F89}
\ee
although his expression identified by $F$ only that portion of the energy flux deposited above the CS (this relation is plotted 
with a dashed line on the right panel of \fig\ \ref{fig:vel_fit}).  The scaling and pre-factor of \eq\ (\ref{eq:F89}) match our expression for {\em evaporation} velocity, \eq\ (\ref{eq:ve}).  Fisher's derivation does not match ours precisely, although it is also possible to find the scaling from dimensional analysis alone.  Curiously, the condensation velocity data, on the right panel of \fig\ \ref{fig:vel_fit}, is well fit by the scaling $F/\rho_{ch,0}$ to a one-half power rather than the one-third power of \eq\ (\ref{eq:F89}) shown as a dotted line.  The discrepancy may result from a systematic variation in the fraction of the total 
flux $F$ reaching the CS.

\subsection{Variation in TR}
\label{sec:tr_var}

\begin{figure}[htp]
\epsscale{1.0}
\plotone{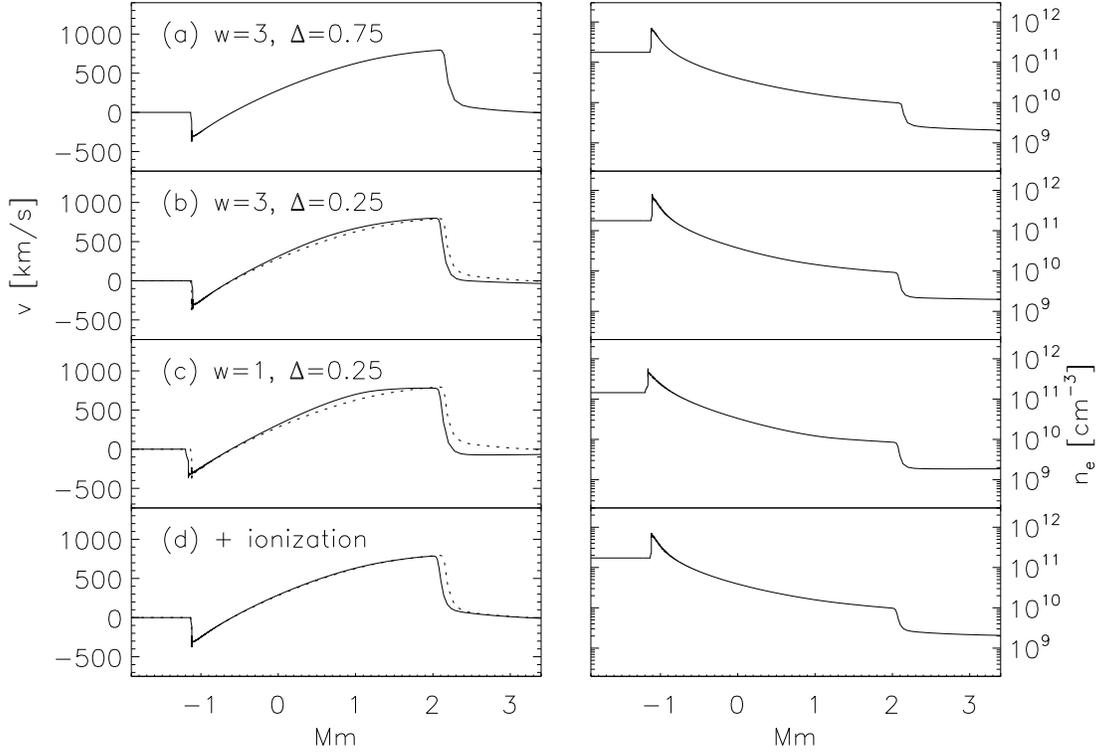}
\caption{Evaporation flows at $t=4$ s from runs with different TR models.  Velocity $v$ in km/s (left column), and the electron density $n_e$ (right) column) are plotted for each run, but the $x$ axis is shifted to line up the plots.
Except for $w$ and $\Delta_{\rm tr}$, all have the same parameters as those in \fig\ \ref{fig:tube_prof}. The top row (a) shows that same run, with $w=3$ Mm and $\Delta_{\rm tr}=0.75$ Mm.  The next two rows have 
$w=3$ Mm and $\Delta_{\rm tr}=0.25$ Mm (b), and $w=1$ Mm and $\Delta_{\rm tr}=0.25$ Mm (c).  The bottom row (d) is the same as (a) but with a specific heat $\cv$ modified to simulate ionization, as explained in the appendix.  Dotted lines in the left column repeat the velocity profile for row (a) for reference.}
	\label{fig:tr_var}
\end{figure}

In order to provide maximum flexibility, a simplified TR model has been used under the
presupposition that the detailed TR structure does not affect the evaporation or condensation.  Figure \ref{fig:tr_var} shows that this presupposition is warranted.  Velocity and density are shown at $t=4$ s from solutions with different TR structures.  The thickness of the TR is varied from $w=1$ Mm to $w=3$ Mm.  The size of the heat sources range from $\Delta_{\rm tr}=0.25$ Mm to $\Delta_{\rm tr}=0.75$ Mm.  It is evident that the resulting flow structure is insensitive to the structure of the initial TR.

\section{Application to more realistic simulations}
\label{sec:5}

The foregoing considered simulations with an artificial TR in order to fully explore the dependence of evaporation on parameters.  In cases with more realistic TRs these parameters cannot be varied independently.  The results derived above do, however, apply to these cases.

\subsection{Simulation with radiation}
\label{sec:rad_run}

We use the same code to perform runs with a more realistic TR by including radiation and gravitational stratification.  The heating function in this case is
\be
  \dot{Q} ~=~ -n_e^2\Lambda(T) ~+~ \heq(\ell) ~+~ H_{\rm fl}(\ell) ~~,
\ee
where $n_e = 0.874\rho/m_p$ is the electron number density assuming complete ionization.  The radiative loss function 
$\Lambda(T)$ is a piece-wise power law fit to the output of {\sc Chianti} 7.1 with coronal abundances
\citep{Landi2013,Dere1997}. We make this function monotonic, $\Lambda\sim T^{1.66}$, over the range $19,000\,{\rm K}<T<78,700\,{\rm K}$.  This omits a peak at $T\simeq20,000$ K, due to silicon which would make the lowest temperature regions susceptible to radiative instabilities.  For $T<19,000$ K we set $\Lambda=0$ in order to prevent any thermal instability in the isothermal chromosphere.  The equilibrium heating, $\heq(\ell)$ is taken to be uniform except in the chromospheric layer, where it is set to exactly balance the radiative losses in equilibrium.  In order to produce a stratified chromosphere we take $g_{\rm par}=\pm 274\,{\rm m/s^2}$ within the chromosphere only.

The initial state is taken to be an equilibrium for heating $\dot{Q}=-n_e^2\Lambda+\heq$ with specified peak temperature 
$T_{co,0}=2\times10^6$ K, a chromosphere at $T_{ch,0}=20,000$ K, and (full) loop length $L=60$ Mm.  Following the arguments of \citet{Rosner1978}, specifying these parameters fixes the equilibrium heating, and coronal pressure, to be $\heq=1.3\times10^{-3}\,{\rm erg\,cm^{-3}\,s^{-1}}$ and $p_0=0.65\,{\rm erg\,cm^{-3}}$.  We add the flare heating profile $H_{\rm fl}$ from \eq\ (\ref{eq:Hfl}), with $F=3.4\times10^{9}$ and $10^{10}\,{\rm erg\,cm^{-2}\,s^{-1}}$ in separate runs.  The characteristics of the resulting evaporation and condensation, plotted as $\times$s in \figs\ \ref{fig:temp_fit} and \ref{fig:vel_fit}, fall very close to those of runs with artificial TRs.  
Notably, the radiative losses decreased the flare temperature (by about 25\% below \eq\ [\ref{eq:Tfl}] in both cases), and the ES density enhancement  (by about 10\% below \eq\ [\ref{eq:Res}] in both cases); radiative losses left the evaporation velocity within 8\% of relation (\ref{eq:ve}).  This corroborates our assumption that radiative losses do not make significant contributions to the rapid evolution of  flare evaporation, at least when it is driven by conduction.

\subsection{Simulations in the literature}

Numerous simulations of conductively driven flaring loops have been reported in the literature.  Each of these used a different code, implementing different aspects of the relevant physics.  Evaporation and condensations characteristics from some notable simulation reports are given in Table \ref{tab:lit} and
plotted as various symbols on \figs\ \ref{fig:temp_fit} and \ref{fig:vel_fit}.  These lie near enough to the dashed lines to be considered to be adequately predicted by relations described in previous section.

\begin{table}[htbp]
{\footnotesize
\begin{tabular}{l|lll | lllllllll}
& & & & & &  \multicolumn{3}{|c|}{ES$^f$} & \\
& $L^a$ & $F^b$ & $p_0^c$ & $t_{\rm shock}^d$ & $T_{\rm fl}^e$& $n_1^g$ & $n_2^h$ & $R_{es}$ & 
$v_e$ & $v_{c}$ \\
& [Mm] & [$10^{10}$] & & [s] & [MK] & \multicolumn{2}{l}{$[10^9{\rm cm^{-3}}]$} 
& & [km/s] & [km/s] \\ \hline
\citet{Cheng1983}  & 10 & 0.2 & 2.9 & 4.95 & $11^j$ & 1.6 & 4.3 & 2.7 & $200$ & 40 \\
\citet{Emslie1985} & 21 & 0.67 & 0.91 & 10 & 20 & 0.64 & 2.9 & 4.6 & 370 & 60 \\
 \multicolumn{1}{r|}{\rule[3pt]{1cm}{0.5pt}} & 21 & 1.3 & 0.91 & 20 & 28 & 0.25 & 1.2 & 4.8 & 670 & 30 \\
\citet{MacNiece1986} & 24 & 0.1 & 2.9& 13.75  & 10 & 1.3 & 2.5 & 2.0 & 60 & 20 \\
\citet{Peres1993},~~r1 & 38 & 0.63& 6.0 & 30 & 22 & 0.65 & 1.5 & 2.3 & 400 & 40 \\
 \multicolumn{1}{r|}{\rule[3pt]{1cm}{0.5pt}~~~r2}  & 38 & 6.3 & 6.0 & --- & 42 & --- & --- & --- & 750 & ---\\
\hline
\end{tabular}\hfil\break
\vskip12pt
\noindent $a$ Full length between chromospheric footpoints.\\
\noindent $b$ Heat flux in units of $10^{10}\,{\rm erg\,cm^{-2}\,s^{-1}}$.\\
\noindent $c$ Pre-flare pressure at base of corona in units of ${\rm erg/cm^{3}}$.\\
\noindent $d$ Time at which evaporation properties were measured.\\
\noindent $e$ Flare temperature at loop apex at $t_{\rm shock}$.\\
\noindent $f$ Properties of the evaporative shock (ES).\\
\noindent $g$ Pre-shock electron density.\\
\noindent $h$ Post-shock electron density.\\
\noindent $j$ Electron temperature.}
\caption{Flare simulations from the literature.}
	\label{tab:lit}
\end{table}

\citet{Peres1993} used the Palermo-Harvard numerical code \citep{Peres1982} to perform several simulations aimed at exploring the effects of viscosity on evaporation.  Two sets of these runs, named run 1 and run 2, use a steady {\em ad hoc} heating source, localized to the loop top.  Theirs had a 10 Mm wide Gaussian profile in place of the flat-topped function of \eq\ (\ref{eq:Hfl}).  The flaring energy flux
\be
  F ~=~ \int\limits_0^{L/2}\, H_{\rm fl}(\ell)\, d\ell ~~,
	\label{eq:Ffl}
\ee
is $F=6.3\times10^9$ (run 1) and $6.3\times10^{10}\,{\rm erg\,cm^{-2}\,s^{-1}}$ (run 2) in their two cases.  Their initial condition was an equilibrium, subject to optically thin radiative losses, and gravity.  The semi-circular loop had a full length of $L=38$ Mm, temperature and pressure $T=3\times10^6$ K, and $p_0=6\,{\rm erg\, cm^{-3}}$, at the loop apex.  Their simulation solved single-fluid hydrodynamic equations with temperature-dependent ionization of hydrogen.

\begin{figure}[htp]
\plottwo{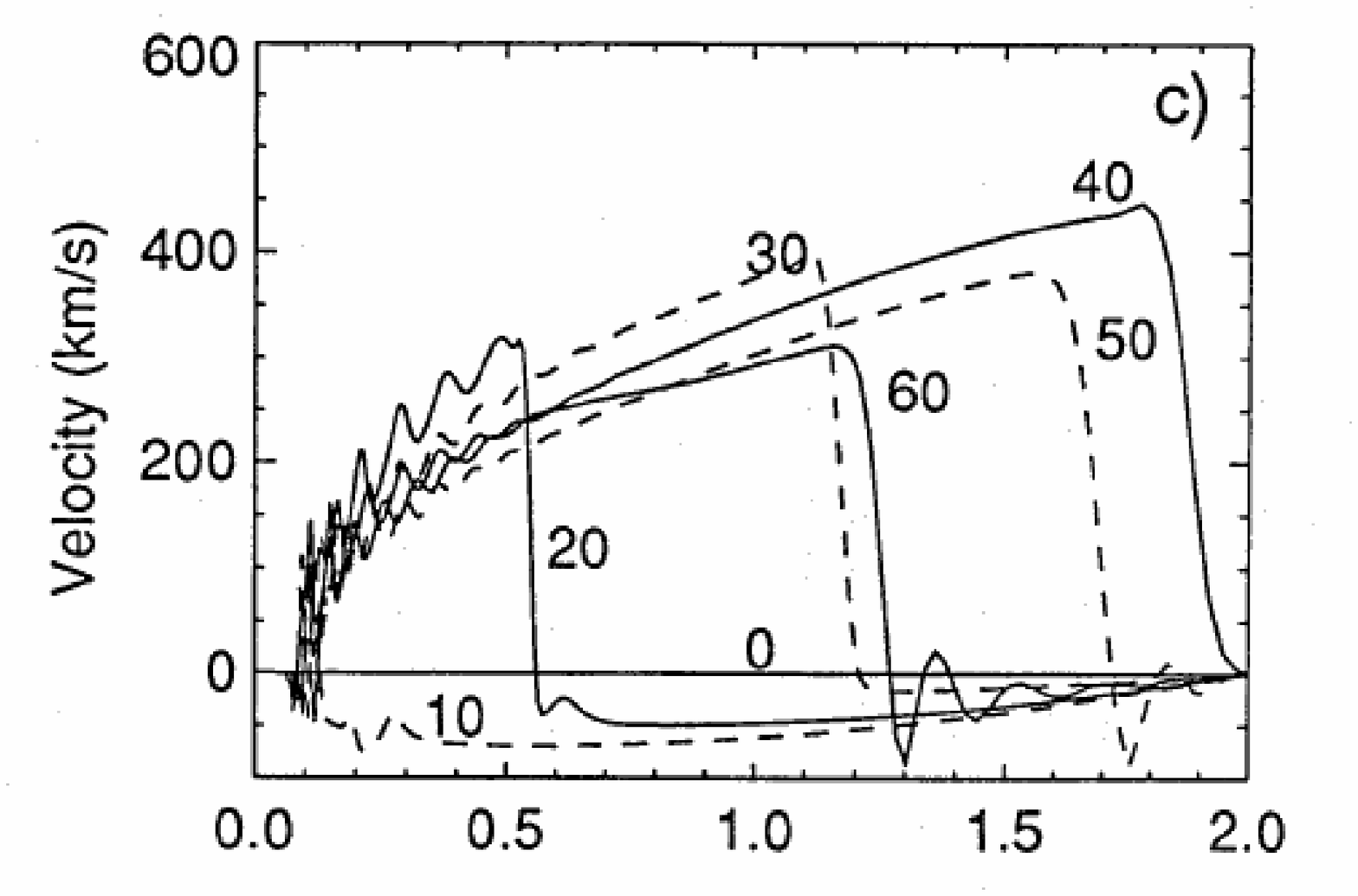}{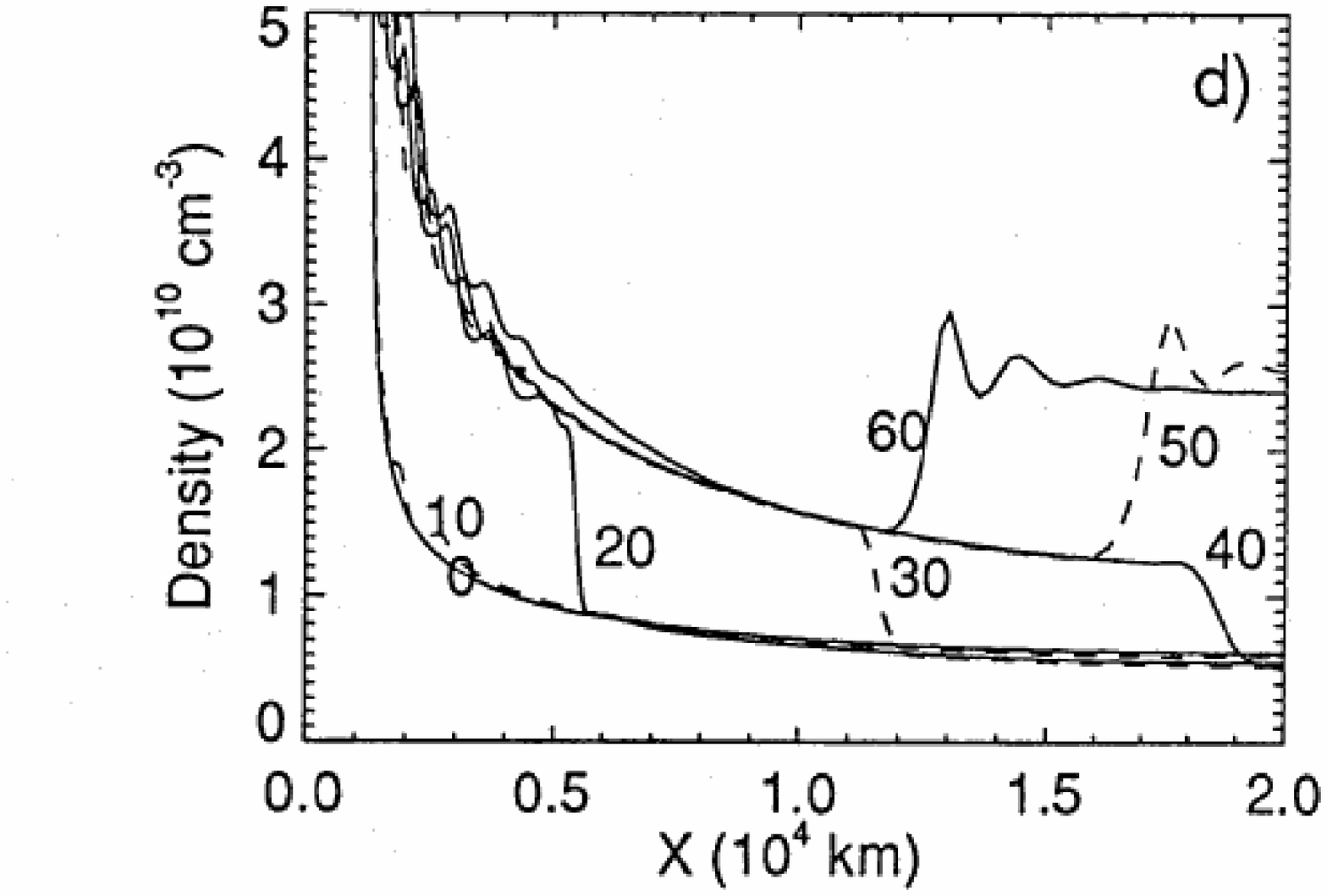}
\caption{The evolution of run 1 with zero viscosity of \citet{Peres1993}, depicted in their \fig\ 1.  
Numbers on each curve give the time in seconds. (Reproduced with permission from Astronomy \& Astrophysics, {\footnotesize\copyright} ESO).}
	\label{fig:Peres}
\end{figure}

Figure \ref{fig:Peres}, reproduced from \citet{Peres1993}, shows the time evolution of run 1 
done with zero viscosity.  The evaporation shock, propagating upward until after $t=40$ sec, is seen clearly due to the low viscosity.  The curves from $t=30$ s are used to deduce properties of the shock, including pre-shock and post-shock density, $n_1$ and $n_2$, and evaporation velocity $v_e$.  The values found are listed in table \ref{tab:lit} and plotted as diamonds in \figs\ \ref{fig:temp_fit} and \ref{fig:vel_fit}.  \citet{Peres1993} report, in a table, 
$T_{\rm fl}=4.2\times10^7$ and $v_e=750$ km/s for run 2.  Since they provide no plot analogous to our \fig\ \ref{fig:Peres} for that run, however, we could not deduce $R_e$ or $v_c$.

One of the earliest simulations of a flaring loop was done by \citet{Cheng1983} using the NRL Dynamic Flux Tube Model 
\citep{Boris1980}.  This code solves quasi-neutral gas dynamics equations for a fully-ionized plasma with distinct electron and ion temperatures.  There is no viscosity.  The electrons are heated directly by an {\em ad hoc} source with a Gaussian profile with full width 2 Mm.  The heating is increased linearly with time so that the energy flux, defined according to \eq\ (\ref{eq:Ffl}), is $F=(4\times10^8\,{\rm erg\,cm^{-2}\,s^{-2}})t$.  The tube is quite short, with full length $L=10$ Mm between the chromospheric feet of the semi-circular loop.  The state at $t=4.95$ sec, shown in their \figs\ 4 and 6, clearly shows an ES structure from which we deduce shock properties.  We list these values in table \ref{tab:lit} and plot them as squares in \figs\ \ref{fig:temp_fit} and \ref{fig:vel_fit}.

\citet{Emslie1985} simulate a flaring loop energized by either  a beam of non-thermal electrons or a direct {\em ad hoc} heating.  They use the code reported in \citet{Nagai1980} solving single-fluid equations with temperature-dependent ionization of hydrogen, viscosity, and a sophisticated treatment of thermal conduction.   The case with 
{\em ad hoc} heating, applicable to our present study, has a loop-top 
Gaussian profile of full width $4.8$ Mm, which ramps up linearly as $F=(6.7\times10^8\,{\rm erg\,cm^{-2}\,s^{-2}})t$.  With a total length $L=22$ Mm, their semi-circular loop is twice as long as that of \citet{Cheng1983}.  From their \figs\ 7 and 8, showing the state of the simulation, we can deduce the properties of the shock at $t=10$ and $20$ s.  These values are table \ref{tab:lit} and plotted as triangles  in \figs\ \ref{fig:temp_fit} and \ref{fig:vel_fit}.

Finally, \citet{MacNiece1986} performed a simulation with very high spatial resolution of a flare heated by a loop-top 
Gaussian profile (7 Mm wide) with constant heat flux $F=10^9\,{\rm erg\,cm^{-2}\,s^{-1}}$.  His \fig\ 7 shows the state of the simulation at a set of times. We use the the latest time shown ($t=13.75$ s) to deduce properties of the evaporation which are plotted as pentagons  in \figs\ \ref{fig:temp_fit} and \ref{fig:vel_fit}.  This very low evaporation velocity appears to follow the departing trend attributed above to a low energy flux.

\section{Discussion}

The foregoing has used simplified numerical experiments to study the process of conduction-driven chromospheric evaporation and condensation.  This led to an analytic model which followed from the simplified scenario of the thermal conduction front leaving the density jump of the initial TR at a uniform temperature.  The analytic solution of this isothermal Riemann problem, depicted in \fig\ \ref{fig:Riemann}, predicts an evaporation shock whose Mach number, and thus density jump, depends only on the temperature jump of the pre-shock TR, $R_{\rm tr}$ --- and depends on this only logarithmically.  The implication of this model is that properties of the evaporation and condensation depend on very few aspects of the flaring loop.
While this analytic model departs in some respects from actual solutions, it still turns out to be useful in inferring the scalings of evaporation and condensation.  A set of runs are used to develop these scaling relations with more accuracy.  These semi-empirical relations are given as \eqs\ (\ref{eq:Tfl}), (\ref{eq:Res}), (\ref{eq:ve}), and (\ref{eq:vc}).

In order to facilitate the exploration of parameter space we adopted a simplified model for the pre-flare TR.  This was motivated by our presupposition that the TR appeared to the flare as a simple density jump.  It was shown in Sec.\ \ref{sec:tr_var} that the specifics of this simplified TR have virtually no effect on the flows.  The scalings were then applied, in Sec.\ \ref{sec:5}, to simulations with more realistic TRs done in this work, and in previous investigations from the literature.  The reasonable level of conformance of these cases to the scaling laws supports our presupposition about the role of the TR.
One source of complication is that density increases with depth in a realistic chromosphere.  In our comparisons to realistic cases we used the local density ahead of the condensation front.  Following this prescription, the condensation velocity, given by \eq\ (\ref{eq:vc}), will decrease with depth in a chromosphere of increasing density.

In a further effort to focus our analysis, we simplified the dynamical equations by excluding aspects deemed inessential to the evaporation.  The momentum equation included viscosity, often reduced below the classical value, but omitted gravity.  The energy equation included classical thermal conductivity, but omitted radiative losses (at least from most runs).  While radiative losses are critical in maintaining the equilibrium structure of the TR, we felt they made less critical contributions to the very rapid dynamics of evaporation.  We tested this hypothesis by performing two runs, reported in Sec.\ \ref{sec:rad_run}, with optically thin radiative losses.  The additional losses reduced both the peak flare both the peak temperature and ES Mach number in comparison to cases with no radiation at all.  The fact that losses reduced these by less than 25\% suggests that radiation was indeed too slow to significantly affect the energy budget of rapid evaporation.

Unlike the thermal conductive cases here considered, a non-thermal electron beam can deposit energy directly into layers of much higher density and lower temperature where radiative time-scales will be shorter.  These losses have been shown to make significant, even dominant, contributions to the evaporation dynamics \citep{Fisher1989b}.  At those temperatures and densities, radiative transfer cannot be treated as optically thin, so a much more sophisticated treatment is required \citep{McClymont1983,Allred2005}.  The most basic energetic effect of such radiative transfer is to permit losses, but prevent them from being instantaneous, as they would be under an optically thin treatment.  This admittedly simplistic argument suggests that  sophisticated radiative treatments of conductively-driven flares should produce evaporation falling somewhere between the cases of no losses and of optically thin losses, that we report.

The relations we report will provide the means of incorporating the effects of chromospheric evaporation into simulations focused on the coronal aspects of a solar flare, including magnetic reconnection.  To develop the relations we adopted the common measure of energizing the loop with an {\em ad hoc} heat source.  We expect, however, that our results will be applicable to cases where the plasma is energized self-consistently, for example, by reconnection-generated shocks.  Future work is aimed at verifying this as well as investigating the effect that evaporation has on the reconnection and its shocks.

\acknowledgements

The author thanks Sean Brannon, George Fisher and Roger Scott for helpful comments and discussion, and thanks the anonymous referee for comments which improved the manuscript. This work was funded by NASA grant NNX13AG09G.

\appendix
\section{Neglected Effect of Ionization}

The foregoing shows how energy input from an {\em ad hoc} source, standing in for flare reconnection, heats and accelerates plasma from the chromosphere.  In order to simplify the model, complete ionization was assumed, even of the chromospheric plasma.  This assumption leads to an over-estimate of the final temperature achieved, $T_*$, since the energy required to achieve full ionization would reduce the amount available for heating and acceleration.  This under-estimate is, however, relatively small and neglecting it leads to a relatively small error.

To estimate the magnitude of the error we consider only that portion of the specific energy, $\varepsilon_*=\cvo T_*$ required by the condensation shock to heat the material from its initial chromospheric state.  Here
$\cvo=(3/2)\kb/\bar{m}$ is the specific heat when fully ionized, and the initial temperature is set to zero since
$T_*\gg T_{ch,0}$.  Had that same specific energy, $\varepsilon_*$, been devoted to both heating 
{\em and} ionizing the plasma, the final temperature achieved would be lower by 
$\DTion=\eion/\cvo$, where $\eion$ is the energy required to fully ionize a unit mass of chromospheric plasma.  The energy required to heat the extra electrons is already accounted for by using $\cvo$ for the specific heat throughout the ionization process.

We estimate the ionization energy using the fairly conservative approach of considering only hydrogen and helium, whose mass fractions we denote $X=0.74$ and $Y=0.25$ respectively, but counting the energy required to ionize both from purely neutral ground states.  The ionization energy under this scenario is
\be
  \eion ~=~ X{\chi_{_{\rm H}}\over m_p} ~+~ Y{\chi_{_{\rm He}} \over 4m_p}~=~ {15.1\,{\rm eV}\over m_p}
\ee
where the energy to fully ionize an atom from its  neutral ground state is 
$\chi_{_{\rm H}} =13.6$ eV for hydrogen and $\chi_{_{\rm He}}=24.6+54.4=79$ eV for helium.  The energy removed by ionization thus lowers the final temperature of the condensation shock by
\be
  \DTion ~=~ {\eion\over \cvo} ~=~ {2\over 3}{\bar{m}\over m_p}\,{15.1\,{\rm eV}\over\kb}
  ~=~ 69,000\,{\rm K} ~~.
  	\label{eq:DT}
\ee
Since the final temperature of the condensation shock is well over $2\times 10^6$ K in the simplified runs, the error from neglecting ionization is less than $5\%$.

To test the foregoing assertion we run a numerical experiment with a specific heat designed to remove energy similar to ionization.  The simulation solves \eqs\ (\ref{eq:cont})--(\ref{eq:erg}) as before except with specific heat
\be
  \cv(T) ~=~ \cvo ~+~ \eion\, f(T) ~=~ \cvo\,\Bigl[ \, 1 ~+~ \DTion\, f(T) \, \Bigr] ~~,
  	\label{eq:cv_mod}
\ee
with $\DTion$ given by \eq\ (\ref{eq:DT}).  The function $f(T)$ expresses the distribution of ionization temperatures and is defined to integrate to unity.  It should peak at temperatures where each specie is being predominately ionized.  Motivated by the description above, however, we use a simple version
\be
  f(T) ~=~ {1\over 2\sigma_T}\,\hbox{sech}^2\left({T-T_0\over\sigma_T}\right) ~~,
\ee
with a single peak of width $\sigma_T=80,000$ K centered at $T_0=250,000$ K.  This removes all energy attributable to ionization, i.e.\ $\eion$, as the plasma is raised from its chromospheric temperature, $T_{ch,0}=20,000$ K to the CS temperature $>10^6$ K.  The function is designed to do this gradually enough to be easily resolved by the simulation.

Aside from replacing \eq\ (\ref{eq:cv}), where $\cv=\cvo$, with the modified form in \eq\ (\ref{eq:cv_mod}), the simulation is exactly the one performed for \fig\ \ref{fig:tube_prof}, with $L=53$ Mm, $F=3.5\times 10^{10}$.  The result, plotted along the bottom row (d) of \fig\ \ref{fig:tr_var} is virtually indistinguishable from the case with $\cv=\cvo$.  To produce any noticeable effect it is necessary to artificially raise the ionization energy, $\eion$, by a factor of ten.  Doing so raises the density in the immediate vicinity of the CS, but otherwise leaves the solution unchanged.  The conclusion is that ionization, had it been legitimately included, would not have had an appreciable effect on the evaporation or condensation flows we have studied here.


\end{document}